\documentclass{aa}  

\usepackage{graphicx}
\usepackage{txfonts}
\usepackage{subcaption}         
\usepackage{lscape}             
\usepackage{placeins}           
\usepackage{threeparttable}

\begin{document}

   \title{Constraining the magnetic field strength of a flaring radio core in the compact steep spectrum  source 3C~138}
\titlerunning{Magnetic field strength in the flaring core of 3C~138}

   \author{Shan Li\inst{1,2}
        \and Sang-Sung Lee\inst{1,2}\corrauth{sslee@kasi.re.kr}
        \and Whee Yeon Cheong\inst{1}
        \and Tao An\inst{3}
        \and Seiji Kameno\inst{4,5,6}
        \and Ruediger Kneissl\inst{4,7}
        }
    \authorrunning{Li et al.}

   \institute{ Korea Astronomy and Space Science Institute, 776 Daedeok-daero, Yuseong-gu, Daejeon 34055, Korea
        \and University of Science and Technology, Korea, 217 Gajeong-ro, Yuseong-gu, Daejeon 34113, Korea
        \and Department of Astronomy, University of Science and Technology of China, Hefei, Anhui 230026, P.R. China
        \and Joint ALMA Observatory, Alonso de Córdova 3107 Vitacura, Santiago 763-0355, Chile
        \and National Astronomical Observatory of Japan, 2-21-1 Osawa, Mitaka, Tokyo 181-8588, Japan 
        \and Department of Astronomy, School of Science, Graduate University for Advanced Studies (SOKENDAI), Tokyo 181-8588, Japan
        \and European Southern Observatory, Alonso de Córdova 3107 Vitacura Casilla 19001 Santiago de Chile, Chile
}
   \date{Received September 30, 20XX}

 
\abstract
{Compact steep spectrum (CSS) sources generally show weak Doppler boosting, yet some exceptions show multi-year-scale radio flux variability and high-energy activity. Since 2022, the CSS quasar 3C~138 has been in a radio high state accompanied by multiple gamma-ray outbursts, offering unique opportunities to study changes in jet physical conditions.}
{We estimated the synchrotron self-absorption (SSA) magnetic field ($B_{\rm SSA}$) in the SSA core of 3C~138 during its high state and compared it with the equipartition magnetic field ($B_{\rm eq}$) to assess the core field environment.}
{Using extended Korean Very long-baseline interferometry Network (KVN) data at 22, 43, 86, and 129~GHz (2024–2025), we calibrated the visibilities and modeled resolved components with circular Gaussians. A single-zone SSA model fitted to the core spectrum provided the turnover frequency and peak flux density, from which we estimated the $B_{\rm SSA}$ and $B_{\rm eq}$. We used Very Large Array and Atacama Large Millimeter/submillimeter Array data to constrain the broadband spectra with the same model.}
{The KVN SSA core shows a turnover at about 33~GHz and a peak flux of about 1.45~Jy. The inferred $B_{\rm SSA}$ is far below equipartition, with $B_{\rm SSA}/B_{\rm eq}\approx0.05$. }
{The flux variability of 3C~138 is driven by a compact, particle-dominated core. Shock-driven particle injection in the inner jet could account for the core brightening and the production of X-ray/gamma-ray emissions through an inverse-Compton process without requiring extreme relativistic beaming effects.}


   \maketitle
   \nolinenumbers

\section{Introduction}
Compact steep spectrum (CSS) sources represent a category of radio-loud active galactic nuclei with projected sizes smaller than 20~kpc and steep spectra ($\alpha<-0.5$, $S_{\nu}\propto\nu^{\alpha}$) above a few hundred megahertz \citep{fanti1989}. They are generally interpreted as young radio sources in early evolutionary stages (e.g., younger than $10^5$ years), though their compactness may also be explained by confinement within a dense interstellar medium with a $N_ {\rm H}$ of approximately $10^{21}$--$10^{24}$~cm$^{-2}$ \citep{o2021}. 

Consistent with weak Doppler boosting, most CSS objects show only modest radio variability (approximately 10\% on approximately 1~year timescales; \citealt{o1998compact}). Yet on multi-year timescales, quasars within the CSS class can exhibit marked flux-density and spectral changes (e.g., 3C~216, B1828+487, 3C~380, and 3C~138; \citealt{torniainen2005long}). A population study has shown that the few known gamma-ray-bright young radio sources are predominantly quasars and occupy a blazar-like locus in the photon-index versus gamma-ray-luminosity plane, favoring a jet origin for the high-energy emission. In these objects, the radio output is dominated by the core and the approaching jet \citep{principe2021gamma}. Taken together, these results point to evolving physical conditions in compact, inner regions of CSS quasars. 

The CSS quasar 3C~138 ($z=0.759$; \citealt{hewitt1989first}) is one such case, showing both radio variability and detected high-energy activity.
3C~138 is a well-known amplitude and polarization calibrator \citep{perley2017accurate}. On kiloparsec scales it shows a highly polarized, one-sided jet extending approximately 400~mas northeastward to a bright hot spot. Very long-baseline interferometry (VLBI) studies have revealed superluminal motion in the inner jet \citep{shen2001superluminal}, strong Faraday rotation, and a nonuniform Faraday screen near the core \citep{cotton2003faraday}, indicating an active nucleus embedded in a complex environment. These properties make 3C~138 a suitable laboratory for testing how the physical conditions within the core (e.g., the magnetic field strength) behave during variability.

Since 2022, 3C~138 has entered an enhanced activity phase across multiple bands. RATAN-600 monitoring shows a steady flux increase at 11–22~GHz, with pronounced rises at 11.2 and 22.3~GHz in 2024–2025 \citep{atel17104}. Very Large Array (VLA) data indicate that the flux density is substantially elevated relative to historical levels, with the enhancement reaching factors of about 2 near 22~GHz, 3 near 33~GHz, and 4 near 45~GHz \citep{perley2017accurate}. Meanwhile, the \textit{Fermi} Large Area Telescope \citep[LAT;][]{Atwood2009} detected gamma-ray flares from 3C~138, accompanied by enhanced X-ray emission \citep{atel17142}, in 2024 October and 2025 May \citep{atel16845,atel17180}. These contemporaneous radio and high-energy changes point to renewed energization in the inner jet and motivated a focused examination of the core’s magnetic field environment.

In this paper we present the first simultaneous VLBI imaging of 3C~138 in its flaring phase at 22, 43, 86, and 129~GHz with the Korean VLBI Network (KVN). From these multifrequency data we identify the synchrotron self-absorption (SSA) core and estimate its magnetic field strength ($B_{\rm SSA}$). We compare $B_{\rm SSA}$ with the equipartition estimate $B_{\rm eq}$ and use the ratio $B_{\rm SSA}/B_{\rm eq}$ to assess whether the core is dominated by particles or magnetic field energy. We then discuss what this inferred energy balance implies for the origin of the radio brightening and its connection to the high-energy activity. The adopted cosmology is $H_{0} = 67.8~{\rm km~s^{-1}~Mpc^{-1}}$, $\Omega_{\rm m} = 0.308$, and $\Omega_{\rm \Lambda} = 0.692$ \citep{wright2006}. At $z = 0.759$, 1~mas corresponds to 7.58~pc, and $1~{\rm mas~yr^{-1}} = 43.5~c$.

\section{Observations and data reduction} \label{sec:obs}
\subsection{KVN observations}
We conducted simultaneous VLBI observations at 22, 43, 86, and 129~GHz on 2024 December 12 and 2025 January 10 with the extended KVN, which comprises the original three 21 m antennas (Yonsei, Ulsan, and Tamna) plus KVN Pyeongchang (KPC). The addition of KPC increased the total number of baselines to six, enabling amplitude self-calibration and improving image fidelity. Each three-hour session consisted of eight 15-minute scans (about 2~h on source) and used right-circular polarization (RCP) only. The data were correlated with DiFX (December 2024) and a GPU-based correlator (January 2025), then calibrated in the Astronomical Image Processing System \citep{greisen2003aips} with a KVN pipeline \citep{hodgson2016automatic}. Interferometric imaging and circular two-dimensional Gaussian model fitting were performed with Difmap \citep{shepherd1997difmap}. For image total flux density, we used 10\% (22/43~GHz), 20\% (86~GHz), and 30\% (129~GHz) uncertainties \citep{lee2016interferometric}. Any potential bias from circular polarization in the RCP-only mode is expected to be negligible compared to these amplitude-scale uncertainties. Uncertainties of the Gaussian model parameters were estimated following \cite{Fomalont1999} and \cite{lee2008global}.

\subsection{Archive data}
To capture total flux variations, including large-scale jet emission, we used VLA data (1–50 GHz) from the observing-support test project TCAL0009 (PI: Lorant Sjouwerman) and 
processed the data with the Common Astronomy Software Applications pipeline, version 6.5.4 \citep{mcmullin2007casa,2022PASP..134k4501C}. We focused on four epochs (2022 March 20, 2023 July 1, 2024 December 20, and 2025 January 19). For 2022 and 2023, we selected the first A-configuration observation from each year, while the 2024 and 2025 data were chosen to be contemporaneous with the two KVN observations.
We also used the Atacama Large Millimeter/submillimeter Array (ALMA) Calibrator Source Catalogue\footnote{\url{https://almascience.eso.org/sc/}} \citep{kneissl2023operational}, which provides well-calibrated flux estimates for standard ALMA calibrators, thereby extending our spectral coverage to the millimeter/submillimeter regime (approximately $97.5$--$343$~GHz).
Complementing our KVN analysis, we used three higher-resolution Very Long Baseline Array (VLBA) epochs at 23.5~GHz (2022 December 14, 2024 June 18, and 2025 June 21) from project UC003 (PI: Phillip Cigan) to trace the flux and structural evolution of the VLBA core and the inner jet.

\section{Results and analysis} 
\subsection{Multifrequency flux and morphology}
\label{sec:flux}
\begin{figure}[!ht]
\centering
  \begin{subfigure}{4.5cm}
    \centering
    \includegraphics[width=4.5cm]{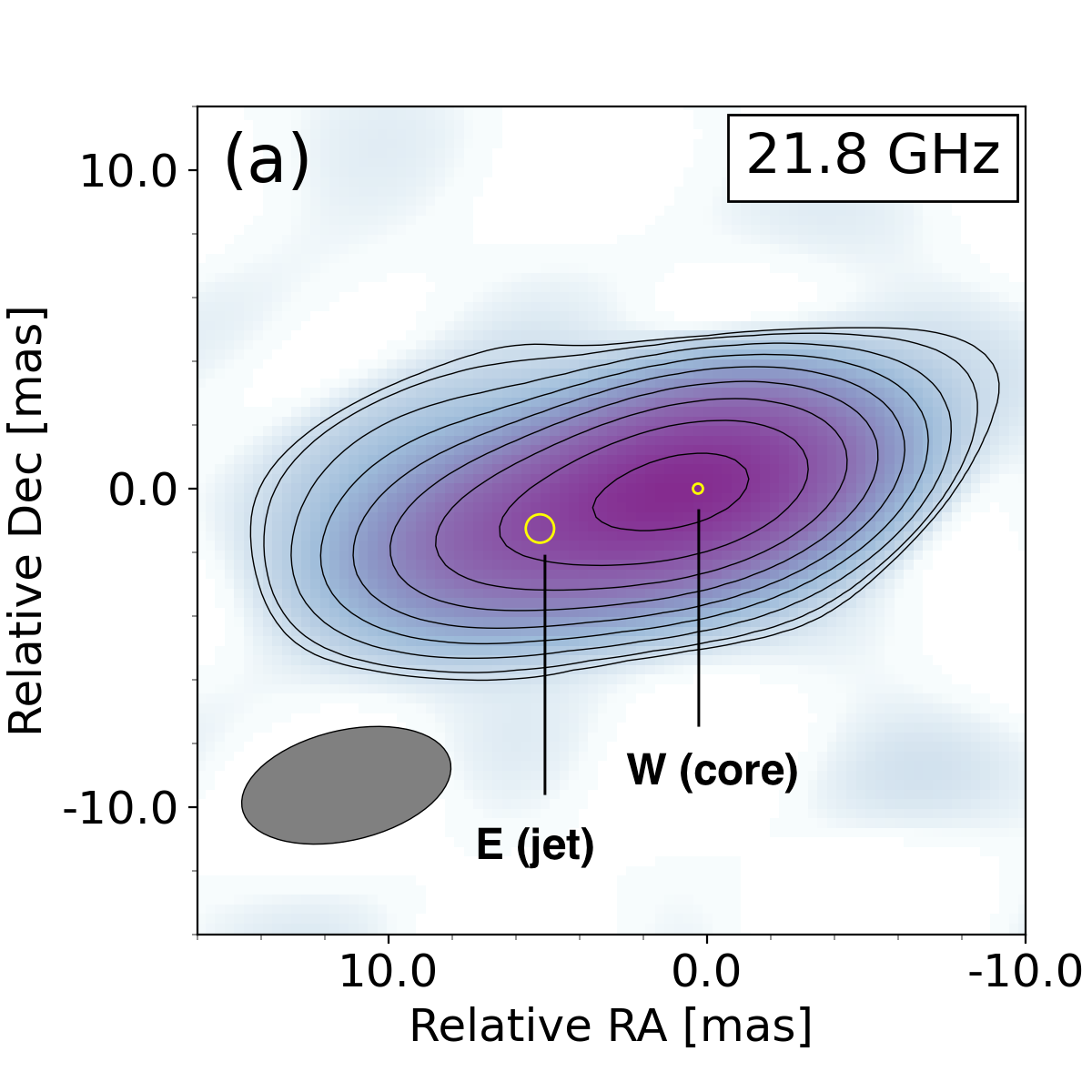}
  \end{subfigure}\hfill
  \begin{subfigure}{4.5cm}
    \centering
    \includegraphics[width=4.5cm]{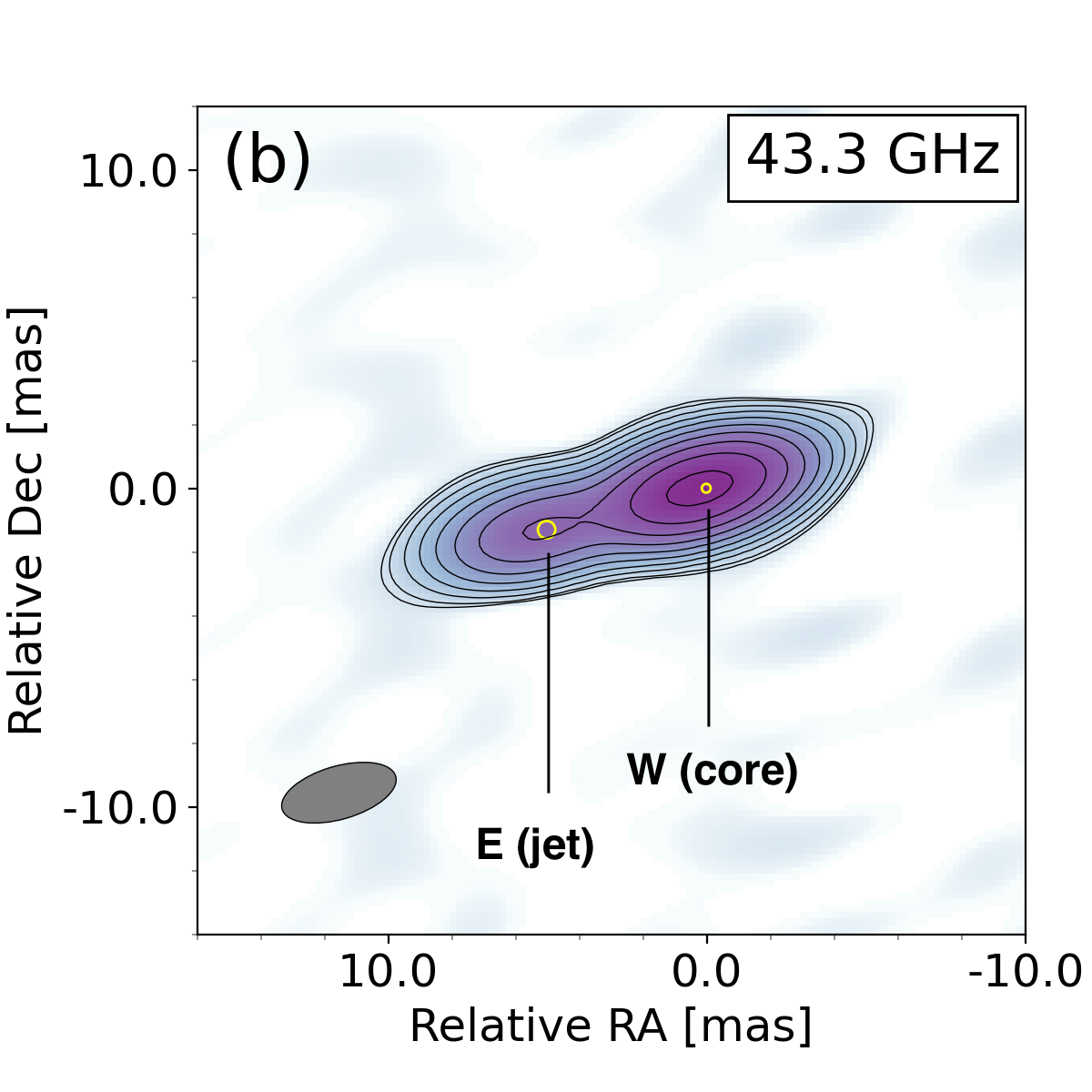}
  \end{subfigure}
  \medskip
  \begin{subfigure}{4.5cm}
    \centering
    \includegraphics[width=4.5cm]{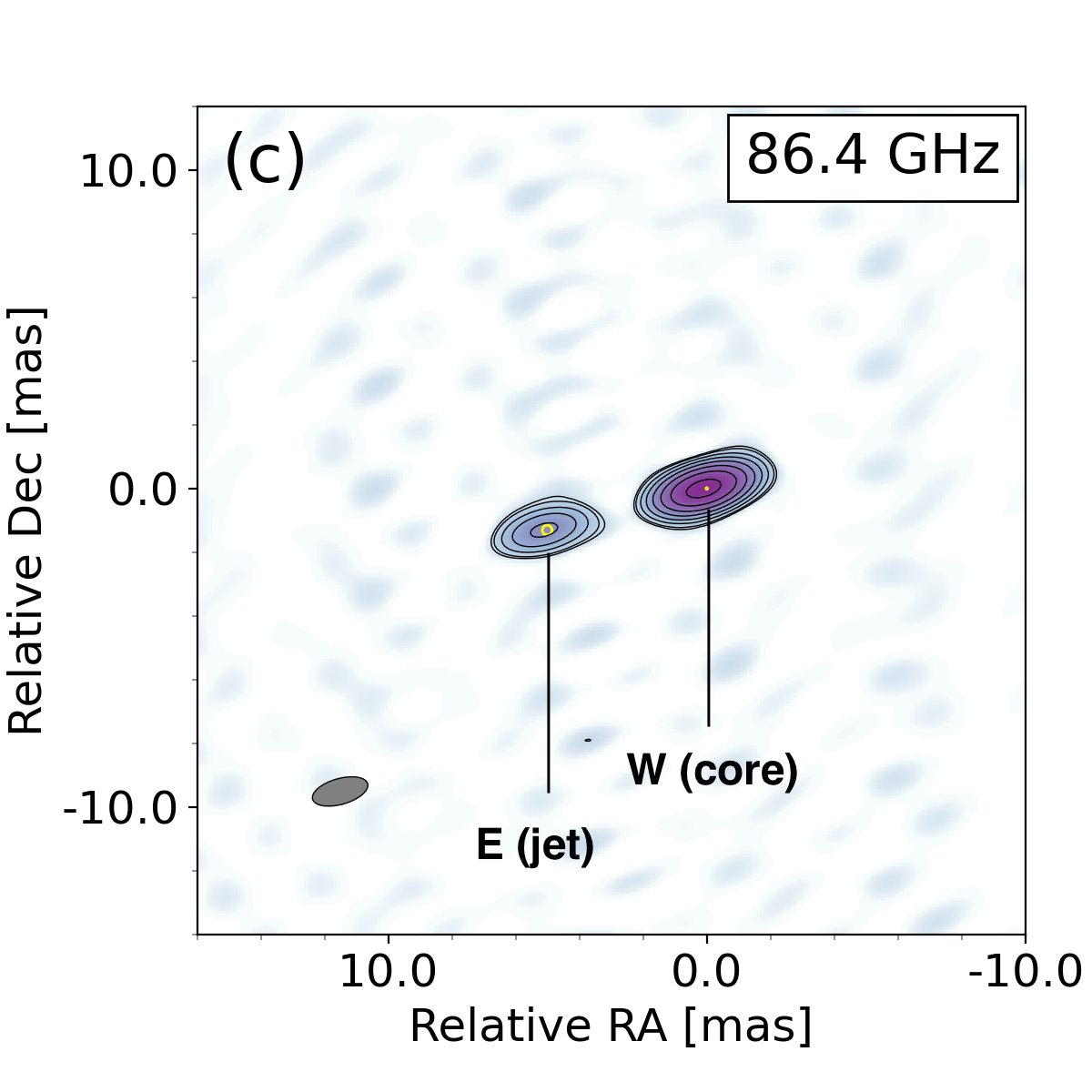}
  \end{subfigure}\hfill
  \begin{subfigure}{4.5cm}
    \centering
    \includegraphics[width=4.5cm]{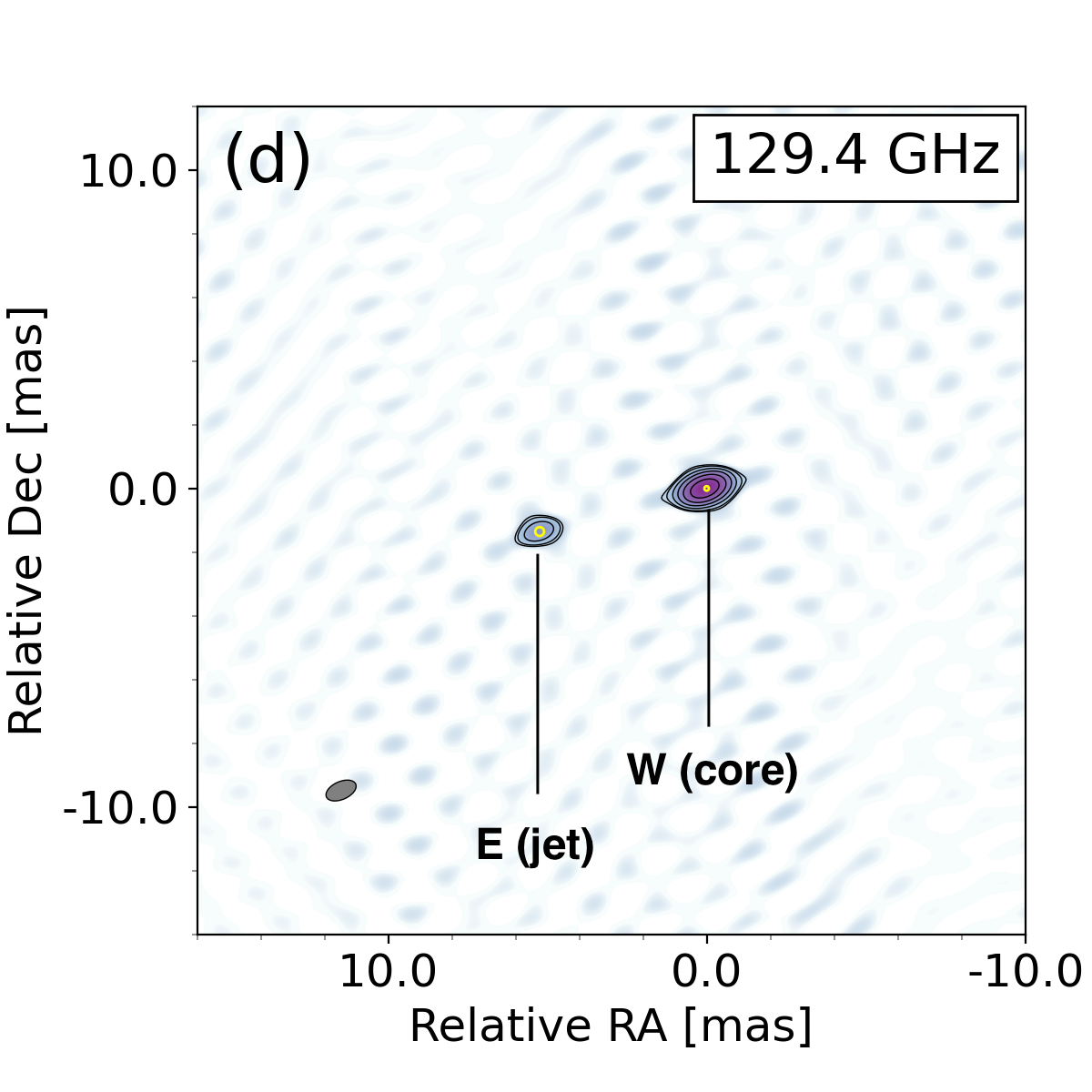}
  \end{subfigure}
\caption{
{KVN multifrequency images of 3C~138 at 21.8, 43.3, 86.4, and 129.4~GHz obtained on 2024 December 12. 
Contours are drawn at $(-1,\,1,\,2,\,4,\,\ldots)\times 5\sigma$, where $\sigma$ is the image rms noise. The synthesized beam is shown in the lower-left corner of each panel. Yellow circles denote circular two-dimensional Gaussian model components. Their diameters represent the fitted full widths at half maximum, and their centers mark the component positions.}
}
\label{fig:kvn_image}
\end{figure}

We detected and imaged the target at 22, 43, and 86~GHz at both epochs. The 129~GHz image is available only for the first epoch due to poor weather. Representative images from 2024 December are shown in Fig.~\ref{fig:kvn_image}.
Over the two epochs, the total CLEAN flux at 22--86~GHz shows at most modest month-scale variability, with no statistically significant change (Table~\ref{tab:kvn_imgparam}). 

Via Gaussian model fitting, we  identified two components separated by about 5~mas, with the fainter eastern component lying to the southeast of the brighter western one at a position angle (PA) of about $105^\circ$ (Table~\ref{tab:kvn_modelfit}). The eastern component is clearly resolved, with fitted sizes larger than the minimum resolvable size, whereas the western component is more compact and is constrained as an upper limit in some bands \citep{lobanov2005resolution}. 
The component flux densities and most fitted component sizes remain consistent between the two epochs within the measurement uncertainties. The larger 86 GHz size of the western component in the second epoch is likely influenced mainly by poorer data quality and calibration effects. The 23.5~GHz VLBA results show two components with a similar separation and PA to those measured with the KVN (Fig.~\ref{fig:vlba_image}). Unlike the eastern component, which remains approximately stable on year timescales, the western component brightened by a factor of about 2.6 from 2022 to 2025 (Table~\ref{tab:vlba_modelfit}). On kiloparsec scales, VLA images reveal a northeastward jet with a PA of approximately $40^\circ$ to $60^\circ$ (Table~\ref{tab:vla_modelfit}), indicating a clear PA offset between the compact two-component structure and the extended jet. As shown in Fig.~\ref{fig:VLA_flux_time}, the VLA core brightens substantially toward the high-frequency bands from 2022 to 2025, while the jet flux varies only weakly. This behavior implies that the observed increase in the integrated VLA flux is dominated by the core component.

\begin{table}[!ht]
\centering
\footnotesize
\caption{KVN image parameters.}
\renewcommand{\arraystretch}{1.25}
\setlength{\tabcolsep}{4pt}
\label{tab:kvn_imgparam}
\begin{threeparttable}
\begin{tabular}{cccccccc}
\hline\hline
 Band & Freq & $S_{\rm CLEAN}$ & $S_{\rm p}$ & $\sigma$ & Beam size & Beam PA \\
  (1) & (2) & (3) & (4) & (5) & (6) & (7) \\
\hline
\multicolumn{7}{c}{2024 December 12} \\  
\hline
K & 21.801 & $1.61 \pm 0.16$ & 1.29 & 1.1 & $6.7\times3.5$ & $-76.9$ \\
Q & 43.346 & $1.70 \pm 0.17$ & 1.37 & 0.6 & $3.7\times1.7$ & $-73.8$ \\
W & 86.436 & $1.15 \pm 0.23$ & 1.03 & 1.7 & $1.8\times0.8$ & $-74.2$ \\
D & 129.398 & $1.11 \pm 0.33$ & 0.90 & 4.5 & $1.0\times0.6$ & $-66.1$ \\
\hline
\multicolumn{7}{c}{2025 January 10} \\
\hline
K & 21.801 & $1.45 \pm 0.14$ & 1.17 & 0.8 & $6.8\times4.2$ & $-70.7$ \\
Q & 43.346 & $1.52 \pm 0.15$ & 1.23 & 1.2 & $3.7\times2.1$ & $-72.2$ \\
 W & 86.436 & $1.14 \pm 0.23$ & 0.95 & 1.7 & $1.8\times0.9$ & $-70.7$ \\
\hline
\end{tabular}
\tablefoot{
Column designations: (1) observing band, (2) central frequency in GHz, (3) total CLEAN flux density with its uncertainty in Jy, (4) peak flux density in Jy\,beam$^{-1}$, (5) image noise level in mJy\,beam$^{-1}$, (6) synthesized beam size in mas $\times$ mas, and (7) beam position angle in degrees. 
}
\end{threeparttable}
\end{table}

\subsection{Spectral analysis}
\label{sec:spectra}
\begin{figure}[!ht] 
\centering 
\includegraphics[width=7cm]{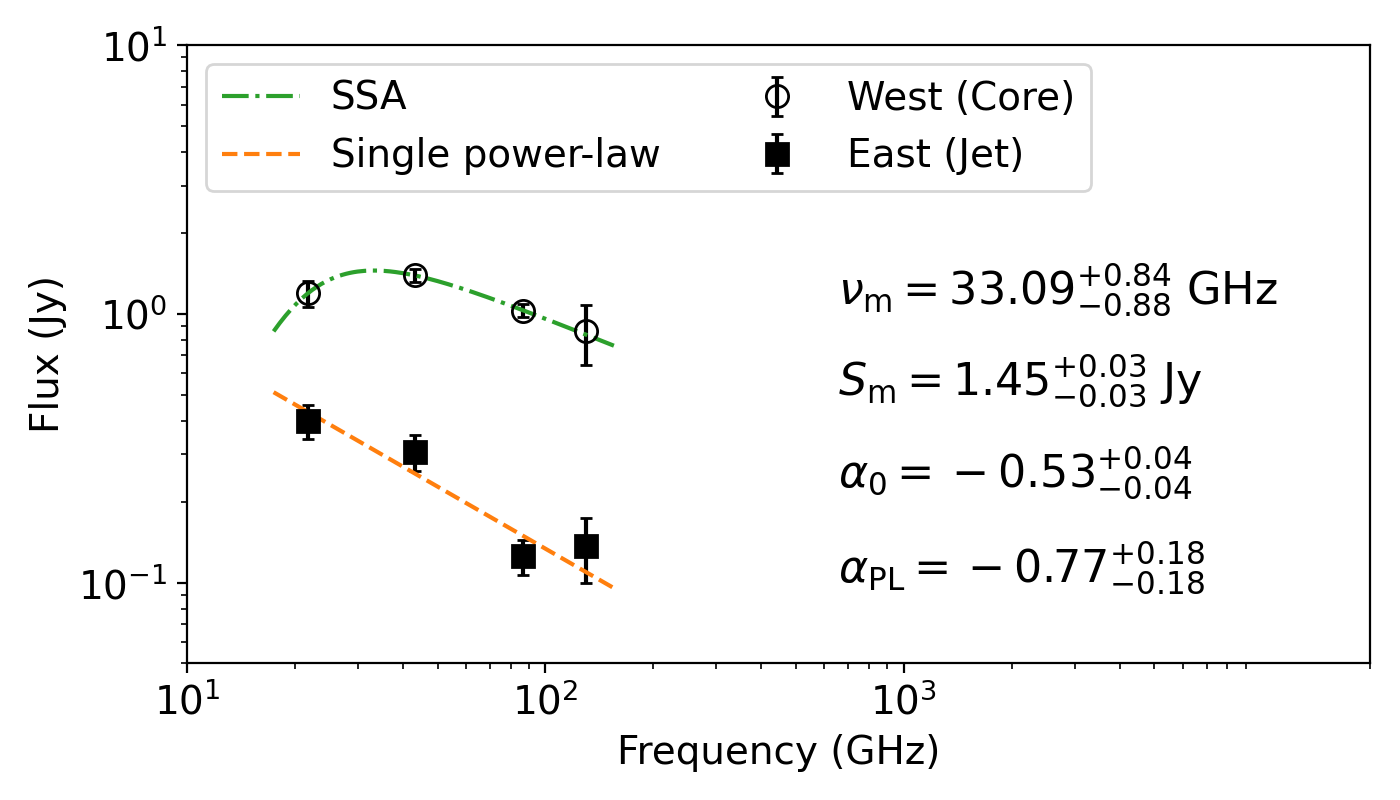} 
\caption{{Spectra of the two KVN-resolved components in 3C~138 on 2024 December 12. Open circles denote the western component (core), fitted with an SSA model (dash-dotted green curve), with turnover frequency $\nu_{\rm m}$, peak flux density $S_{\rm m}$, and optically thin spectral index $\alpha_{0}$. Filled squares denote the eastern component (jet knot), fitted with a PL model (dashed orange line) with spectral index $\alpha_{\rm PL}$. Error bars indicate 1$\sigma$ uncertainties.}} 
\label{fig:KVN_spectrum} 
\end{figure}

The simultaneous four-band KVN observations enabled us to determine the spectra of the two components and identify the core. Figure~\ref{fig:KVN_spectrum} presents the spectra of the first epoch. The western component exhibits an apparently flatter spectrum than the eastern one, with an inverted slope between 22 and 43~GHz, which is characteristic of an SSA core. We therefore fitted the eastern component with a single power law (PL), $S_\nu \propto \nu^{\alpha}$, where $\alpha$ is the spectral index, and modeled the western component with a single-zone SSA model \citep{turler2000modelling}:
\begin{equation}
S_\nu = S_{\rm m} \left(\frac{\nu}{\nu_{\rm m}}\right)^{\alpha_{\rm t}}
\frac{1-\exp\left[-\,\tau_{\rm m}\left(\nu /\nu_{\rm m}\right)^{\alpha_{\rm t}-\alpha_0}\right]}
{1-\rm \exp\left(-\tau_m\right)}\,,
\end{equation}
where $\alpha_{\rm t}$ and $\alpha_0$ are the thick and thin spectral indices, and $\nu_{\rm m}$ and $S_{\rm m}$ are the SSA turnover frequency and peak flux density. 
The spectral fitting results for the two components are shown in Fig.~\ref{fig:KVN_spectrum}.

To probe the time variability of the broadband spectrum, we interpolated the ALMA flux densities using Gaussian-process (GP) regression, paired them with the four VLA epochs, and fitted the combined VLA+ALMA spectra with both a PL model and a PL+SSA model for each epoch. We then selected the preferred model by comparing the reduced $\chi^2$ and the Bayesian information criterion. As the two arrays are not simultaneous, and have unmatched angular resolutions, these fits should be interpreted with caution. The best-fit parameters are presented in Table~\ref{tab:plssa_fits}, and the fitting results are plotted in Fig.~\ref{fig:vla-alma_fitting}.

\section{Discussion}
\label{sec:discussion}
\subsection{SSA core spectrum and temporal evolution}\label{sec:KVN_core_ssa}
The KVN western component shows a flat--inverted spectrum between 22 and 43~GHz and is well described by an SSA model, supporting its identification as the SSA core. The eastern feature follows a single steep PL, consistent with an optically thin inner jet component (Fig.~\ref{fig:KVN_spectrum}).

For the VLA data, as shown in Fig.~\ref{fig:vla-alma_fitting}, a spectral inflection is detected only in 2024 and 2025, whereas the 2022--2023 spectra are adequately described by a single steep PL. This indicates a transition from a PL-only state to a PL+SSA state. Within the PL+SSA state, the optically thin slope ($\alpha_0$) of the SSA component further steepens, suggesting the high-energy electron population is evolving. If driven by a propagating shock, one would expect enhanced millimeter polarization, systematic electric vector position angle (EVPA) evolution, and higher-frequency light-curve peaks to precede those at lower frequencies.

Because the VLA flux increase is core-dominated (Sect.~\ref{sec:flux}) and the VLA core blends the two KVN components, the integrated VLA+ALMA behavior can be used as a qualitative proxy for the SSA evolution of the KVN core. The turnover frequency inferred from VLA+ALMA is systematically higher than that from the KVN core, which may reflect residual mixing with optically thin emission and/or intrinsic inhomogeneity within the KVN core region, such that a one-zone SSA model yields an effective turnover frequency.

\subsection{Magnetic field strength}\label{sec:KVN_core_B}
We estimated the SSA magnetic field strength and the equipartition magnetic field strength of the KVN core for the first epoch (2024 December 12) using the SSA spectral parameters listed in Fig.~\ref{fig:KVN_spectrum}, together with the SSA region size ($d_{\rm m}$) scaled from the bright-state VLBA core size and the Doppler factor ($\delta$) estimated from VLBA core variability (Appendix~\ref{app:Bfield}). 

We adopted $d_{\rm m}=0.07\pm0.01$~mas and $\delta_{\rm var}=2.36^{+0.35}_{-0.38}$, noting that $\delta_{\rm var}$ is an approximate constraint and likely a lower limit given the sparse temporal sampling and possible unresolved substructure. Using these parameters, we derived $B_{\rm SSA}=10.12^{+9.59}_{-5.11}$~mG and $B_{\rm eq}=200.77^{+44.08}_{-33.79}$~mG, which yields $B_{\rm SSA}/B_{\rm eq}=0.05^{+0.06}_{-0.03}\ll 1$. This indicates that the SSA core is particle-dominated.

The dominant uncertainty arises from $d_{\rm m}$ (scaled from the VLBA core size). We propagated uncertainties via Monte Carlo sampling, and we further tested the sensitivity of the result to the assumed geometric and opacity scalings by varying jet-opening index and the core-shift index within plausible ranges. In all cases, we obtain $B_{\rm SSA}/B_{\rm eq}\ll 1$ (Fig.~\ref{fig:B_dependence}), so the particle-dominated conclusion is robust.

\subsection{Core brightening and high-energy emission}
Multifrequency, multi-scale data show that the flux rise since 2022 is core-dominated (Fig.~\ref{fig:VLA_flux_time}), meaning a kiloparsec-scale jet--interstellar medium interaction is likely not the primary driver of the integrated flux increase. VLBA images show a steady brightening of the western core, while the eastern knot and the kiloparsec-scale VLA jet remain nearly constant, pointing to processes within the VLBI core.

The KVN SSA core magnetic field strength is far below the equipartition value, implying a particle-dominated region and an increase in synchrotron emitting electrons, possibly associated with shock-driven particle acceleration. ALMA polarization monitoring further shows a gradual increase in polarized flux after 2022 \citep{kameno20253c}, consistent with a more ordered magnetic field.

If a shock is responsible, a new component may be forming but still unresolved. Using the inverted minimum-resolvable-size relation, we find that the image rms required to resolve a separation of two KVN core sizes is 5.71, 19.1, 2.10, and 17.3~mJy at 22, 43, 86, and 129~GHz in the first epoch, and 0.8, 1.2, and 1.7~mJy in the second epoch at 22, 43, and 86~GHz, which is comparable to or higher than our achieved rms at each frequency at both epochs. Yet no new component is detected near the core, suggesting that any emerging feature remains blended with the core and has not reached a resolvable downstream separation.

Using the inferred $\delta_{\rm var}$, we estimate that the corresponding apparent speed is at most of order $2c$, corresponding to a proper motion of about $0.05$~mas~yr$^{-1}$. Under a separability criterion of two VLBA core sizes (about $0.12$~mas), a newly emerging component would therefore require at least a few years to become clearly separated from the core. If the viewing angle is instead as large as $\theta=34^\circ$, as inferred by \citet{shen2001superluminal}, then the corresponding separation timescale would be even longer.

Either case implies only mild Doppler boosting. Nevertheless, with efficient acceleration, inverse-Compton processes can produce X-ray to gamma-ray radiation under modest Doppler boosting, as suggested by \textit{Fermi}-LAT detections of young radio galaxies and compact symmetric objects \citep[e.g., PKS~1718-649 and NGC~3894;][]{Migliori2016, principe2020ngc3894, principe2021gamma}.
Continued high-frequency, high-resolution polarization VLBI observations will allow us to test this scenario by constraining the emergence timescale of a new component and by tracking polarization and EVPA evolution in the core.

\section{Conclusion}
\label{sec:conclusion}
Using two-epoch, four-band KVN imaging of the CSS source 3C~138, we identified the western of the two KVN-resolved components as the SSA core. The turnover frequency inferred from the integrated VLA+ALMA spectra is higher than that from the KVN core spectrum, which can be attributed to resolution blending of the compact core with additional emitting regions in the lower-resolution data.
The multi-year radio brightening is core-dominated and the core magnetic field is below equipartition, implying a particle-dominated emitting region with only mild Doppler boosting. Shock-driven particle injection in the inner jet could account for the contemporaneous X-/gamma-ray emission without extreme beaming. Although no new knot is resolved, a slowly emerging feature may still be blended within the VLBI core. Continued high-frequency, high-resolution VLBI monitoring will be needed to test this scenario.

\begin{acknowledgements}
We are grateful to the staff of the KVN who helped to operate the array and to correlate the data. The KVN and a high-performance computing cluster are facilities operated by the KASI (Korea Astronomy and Space Science Institute). The KVN observations and correlations are supported through the high-speed network connections among the KVN sites provided by the KREONET (Korea Research Environment Open NETwork), which is managed and operated by the KISTI (Korea Institute of Science and Technology Information). This work is supported by the National Research Foundation of Korea (NRF) grant funded by the Korea government (MSIT) (2020R1A2C2009003, RS-2025-00562700). The authors acknowledge use of the Very Long Baseline Array under the US Naval Observatory's time allocation. This work supports USNO's ongoing research into the celestial reference frame and geodesy. This work also made use of data from the Very Large Array operated by the National Radio Astronomy Observatory (NRAO). The NRAO and the Green Bank Observatory are facilities of the U.S. National Science Foundation operated under cooperative agreement by Associated Universities, Inc. This work makes use of ALMA data (ADS/JAO.ALMA\#2011.0.00001.CAL). ALMA is a partnership of ESO, NSF (USA), and NINS (Japan), together with NRC (Canada), NSTC/ASIAA (Taiwan), and KASI (Republic of Korea), in cooperation with the Republic of Chile. The Joint ALMA Observatory is operated by ESO, AUI/NRAO, and NAOJ.
\end{acknowledgements}

\bibliographystyle{aa} 
\bibliography{ref} 

\begin{appendix}
\nolinenumbers
\section{Magnetic field strength estimates}\label{app:Bfield}
This appendix provides the details of the magnetic-field calculations presented in Sect.~\ref{sec:KVN_core_B}.

\subsection{Size scaling for the SSA region}\label{app:size_scaling}
The 23.5~GHz VLBA core appears larger in the fainter epoch (2022) but becomes much smaller in the brighter epochs (2024--2025; Table~\ref{tab:vlba_modelfit}). This behavior is consistent with the faint-state core being effectively broadened by blending of multiple sub-components within the finite VLBA resolution, whereas the bright-state emission is dominated by a more compact region. We therefore adopt the 2025 June VLBA core size ($d_{\rm VLBA}=0.06\pm0.01$~mas), measured in a bright-state epoch close in time to the KVN observations and with a well-constrained model fit (e.g., a fitted size exceeding the minimum resolvable size), as the reference size to scale the SSA region angular size $d_{\rm m}$, following
\begin{equation}
d_{\rm m}
= 1.6\,d_{\rm VLBA}
\left(\frac{\nu_{\rm m}}{\nu_{\rm VLBA}}\right)^{-\epsilon/k_{\rm r}}~[\rm mas]\,,
\label{eq:dm_app}
\end{equation}
where the factor $1.6$ converts the full width at half maximum of a circular Gaussian component to the equivalent angular diameter of a uniform disk \citep{pearson1995}, $\epsilon$ describes the jet opening ($\epsilon\simeq1$ for a conical jet and $\epsilon<1$ for a collimating flow), and $k_{\rm r}$ is the index that characterizes the frequency dependence of the core position, $r_{\rm core}\propto \nu^{-1/k_{\rm r}}$ \citep{lobanov1998ultracompact}. We adopt $k_{\rm r}=1$ as a commonly used empirical value in core-shift scaling \citep{lobanov1998ultracompact}, and the median $\epsilon$ value of 0.97 from \cite{algaba2017resolving}.

\subsection{Variability Doppler factor}\label{app:doppler}
\begin{figure}[!ht]
    \centering
    \includegraphics[width=8cm]{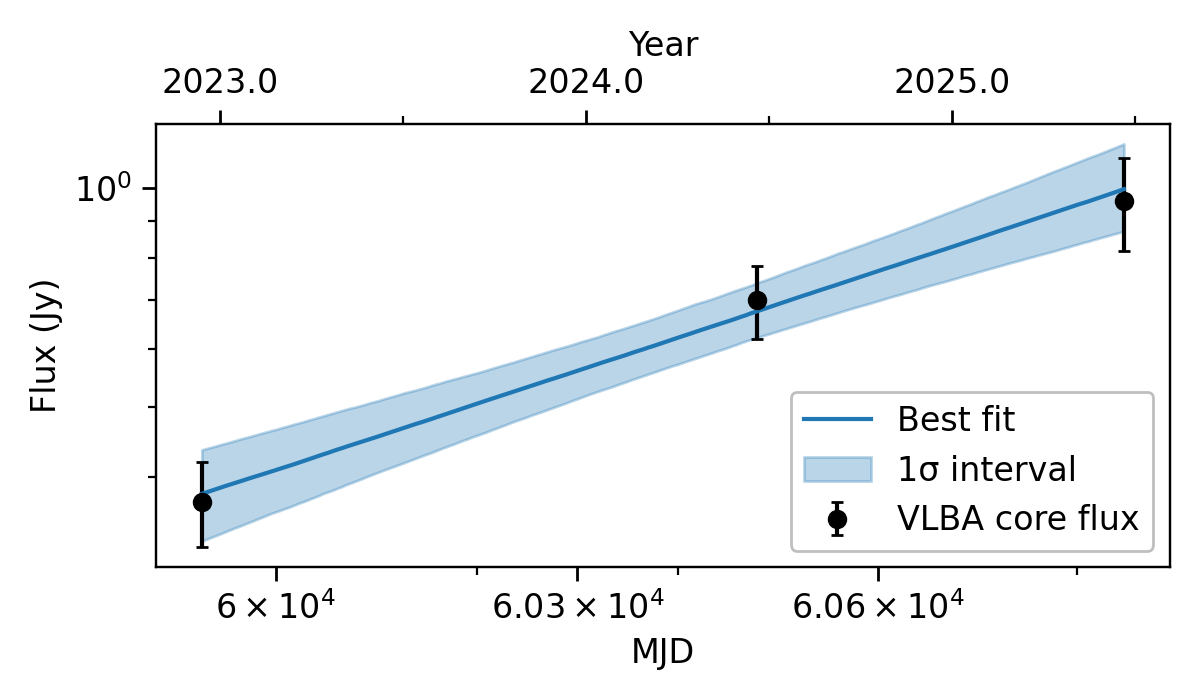}
    \caption{VLBA core light curve at 23.5~GHz from 2022 to 2025. The solid line is a least-squares fit, and the shaded region marks the $1\sigma$ confidence band. Black points show the measured core flux densities with $1\sigma$ errors.}
    \label{fig:tau_var}
\end{figure}

We estimate the variability Doppler factor from the temporal evolution of the VLBA core flux density $S(t)$ at 23.5~GHz. Following Eq.~(\ref{eq:tau_var}), we fit the VLBA core light curve in log space with a least-squares method,
\begin{equation}
\ln S = \ln S_0 + \frac{t-t_0}{\tau_{\rm var}}\,,
\label{eq:tau_var}
\end{equation}
where the fitted slope equals $1/\tau_{\rm var}$ (Fig.~\ref{fig:tau_var}), and $\tau_{\rm var}$ is the variability timescale in days.

We then estimate the observed variability brightness temperature $T_{\rm B}^{\rm var}$, assuming that the variability is dominated by a circular Gaussian emitting region \citep{kang2021}:
\begin{equation}
T_{\rm B}^{\rm var} = 4.077 \times 10^{13}
\left(\frac{D_{\rm L}}{\nu\,\tau_{\rm var}}\right)^2
\frac{\Delta S}{(1+z)^4}~{\rm K}\,,
\label{eq:tb_var}
\end{equation}
where $z$ is the redshift, $D_{\rm L}$ is the luminosity distance in Mpc, $\nu$ is the observing frequency in GHz, and $\Delta S$ (Jy) is the flux density difference between the peak and the value at peak/$e$. 

Assuming an equipartition brightness temperature ($T_{\rm B,eq}$) as the maximum intrinsic brightness temperature, the variability Doppler factor can be estimated \citep[e.g.,][]{fuhrmann2008,hovatta2009,kang2021} as
\begin{equation}
\delta_{\rm var} = (1+z)\left(\frac{T_{\rm B}^{\rm var}}{T_{\rm B,eq}}\right)^{1/3}\,.
\label{eq:delta_app}
\end{equation}
Here we adopt $T_{\rm B,eq}=5\times10^{10}$~K \citep{hovatta2009}. 

Given the sparse sampling of the VLBA core flux density (three epochs) and the possibility of unresolved substructure within the VLBA core, the derived $\tau_{\rm var}$ should be regarded as an upper limit. Consequently, $T_{\rm B}^{\rm var}$ and $\delta_{\rm var}$ are likely underestimated, and $\delta_{\rm var}$ should be treated as an approximate constraint and potentially a lower limit.

\subsection{Magnetic field strength formulae}\label{app:B_formulae}
Given $S_{\rm m}$, $\nu_{\rm m}$, and $\alpha_0$ from the SSA fit (Fig.~\ref{fig:KVN_spectrum}), together with $d_{\rm m}$ and $\delta_{\rm var}$, we compute the SSA magnetic field strength from \citet{marscher1983accurate}:
\begin{equation}
B_{\rm SSA}=10^{-2}\,b(\alpha_0)\,
\left(\frac{S_{\rm m}}{\rm Jy}\right)^{-2}
\left(\frac{d_{\rm m}}{\rm mas}\right)^{4}
\left(\frac{\nu_{\rm m}}{\rm GHz}\right)^{5}
\left(\frac{\delta}{1+z}\right)^{-1}~[{\rm mG}]\,,
\label{eq:Bssa_app}
\end{equation}
where $b(\alpha_0)$ is a factor that depends on the optically thin spectral index, and $\delta$ is the Doppler factor.

The equipartition magnetic field strength is estimated following \citet{kataoka2005x}:
\begin{align}
B_{\rm eq} &= 1.23\times10^{-4}\,
\eta^{2/7}(1+z)^{11/7}
\left(\frac{D_{\rm L}}{100~{\rm Mpc}}\right)^{-2/7} \notag\\
&\quad\times
\left(\frac{\nu_{\rm m}}{5~{\rm GHz}}\right)^{1/7}
\left(\frac{S_{\rm m}}{100~{\rm mJy}}\right)^{2/7}
\left(\frac{10^3 d_{\rm m}}{0.3~{\rm arcsec}}\right)^{-6/7}
\delta^{-5/7}~{\rm [G]}\,,
\label{eq:Beq_app}
\end{align}
assuming $\eta=1$ (the ratio of proton + electron to electron energy densities). 

The inferred $B_{\rm SSA}$ is proportional to $\delta^{-1}$, $B_{\rm eq}$ to $\delta^{-5/7}$, and thus $B_{\rm SSA}/B_{\rm eq}$ to $\delta^{-2/7}$. Since $\delta_{\rm var}$ is likely underestimated, the derived $B_{\rm SSA}$, $B_{\rm eq}$, and $B_{\rm SSA}/B_{\rm eq}$ should be interpreted as upper limits, subject to a corresponding systematic uncertainty.

\subsection{Uncertainty propagation and sensitivity tests}\label{app:uncertainty}
The dominant uncertainty in the magnetic field estimates arises from the SSA region angular size $d_{\rm m}$, which is scaled from the VLBA core size $d_{\rm VLBA}$ (Eq.~\ref{eq:dm_app}). The VLBA core size $d_{\rm VLBA}$ at 23.5~GHz is obtained from circular Gaussian model fitting, and the statistical uncertainties of the fitted parameters are estimated from the post-fit image rms and the component signal-to-noise ratio using the first order approximation of \citet{Fomalont1999}.

\begin{figure}[!ht]
  \centering
    \includegraphics[width=4.4cm]{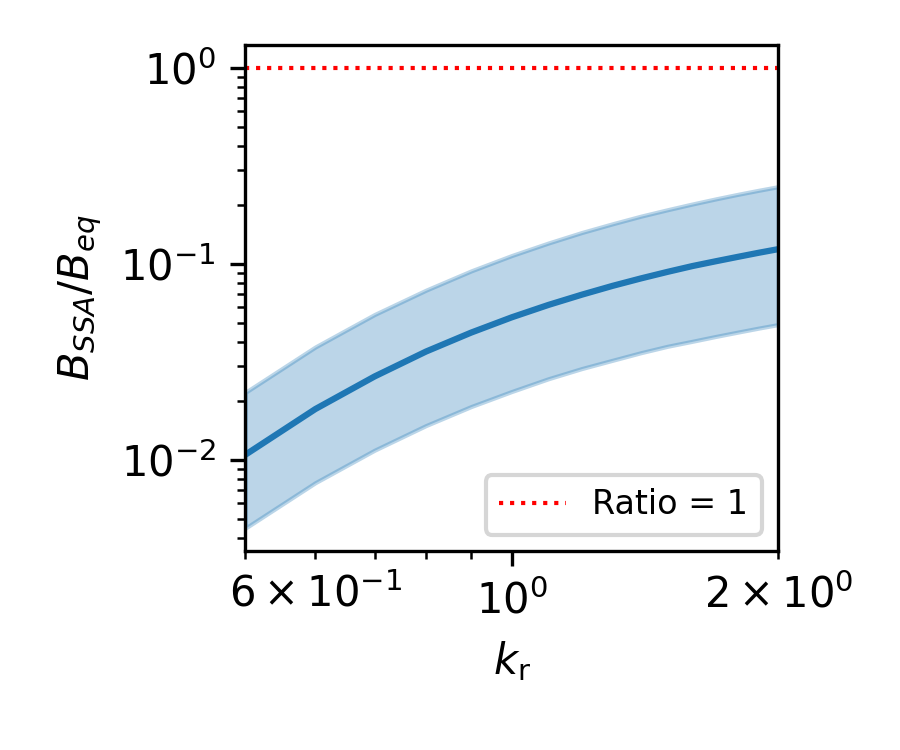}
    \includegraphics[width=4.4cm]{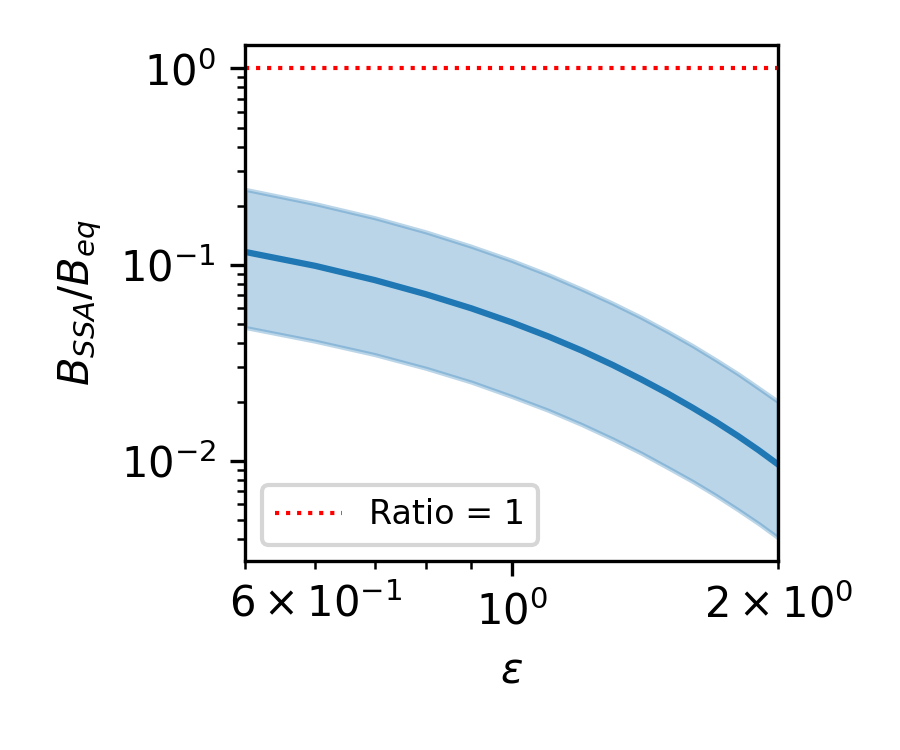}
  \caption{Left: Dependence of $B_{\rm SSA}/B_{\rm eq}$ on the core-shift index ($k_{\rm r}$). Right: Dependence of $B_{\rm SSA}/B_{\rm eq}$ on the jet opening ($\epsilon$). }
  \label{fig:B_dependence}
\end{figure}

To account for the strong non-linearity of Eqs.~(\ref{eq:dm_app})--(\ref{eq:Beq_app}), we propagated uncertainties via Monte Carlo sampling by drawing $S_{\rm m}$, $\nu_{\rm m}$ and $\alpha_0$ from the SSA-fit posterior and $d_{\rm VLBA}$ from its inferred distribution. 
In addition, we explored the impact of the geometric and opacity scalings in Eq.~(\ref{eq:dm_app}) by varying the jet-opening index ($\epsilon$) and the core-shift index ($k_{\rm r}$) within plausible ranges (Fig.~\ref{fig:B_dependence}). 

\section{Additional images and plots}
\label{sec:plots}

\begin{figure}[!ht]
\centering
  \begin{subfigure}{4.5cm}
    \centering
    \includegraphics[width=4.5cm]{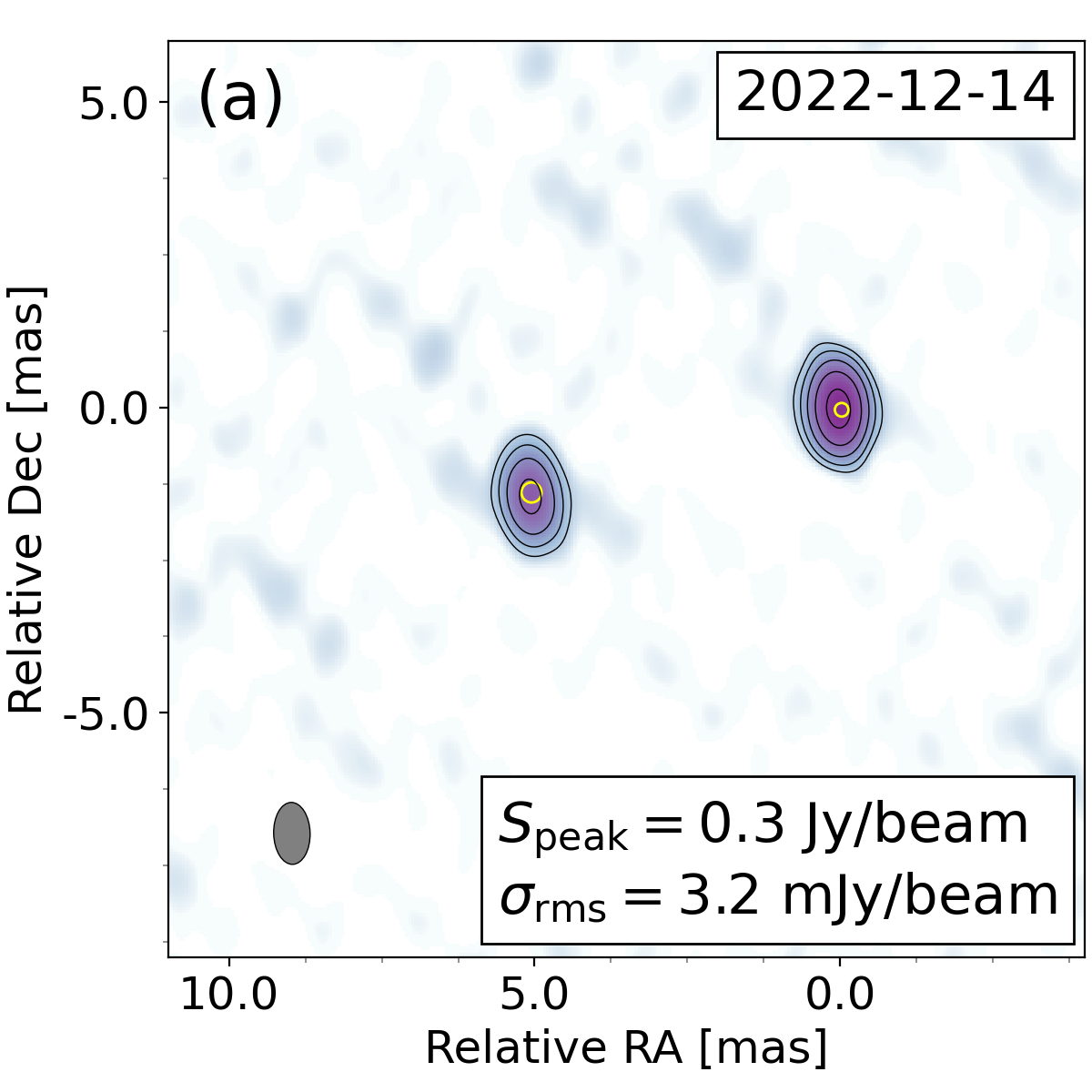}
  \end{subfigure}\hfill
  \begin{subfigure}{4.5cm}
    \centering
    \includegraphics[width=4.5cm]{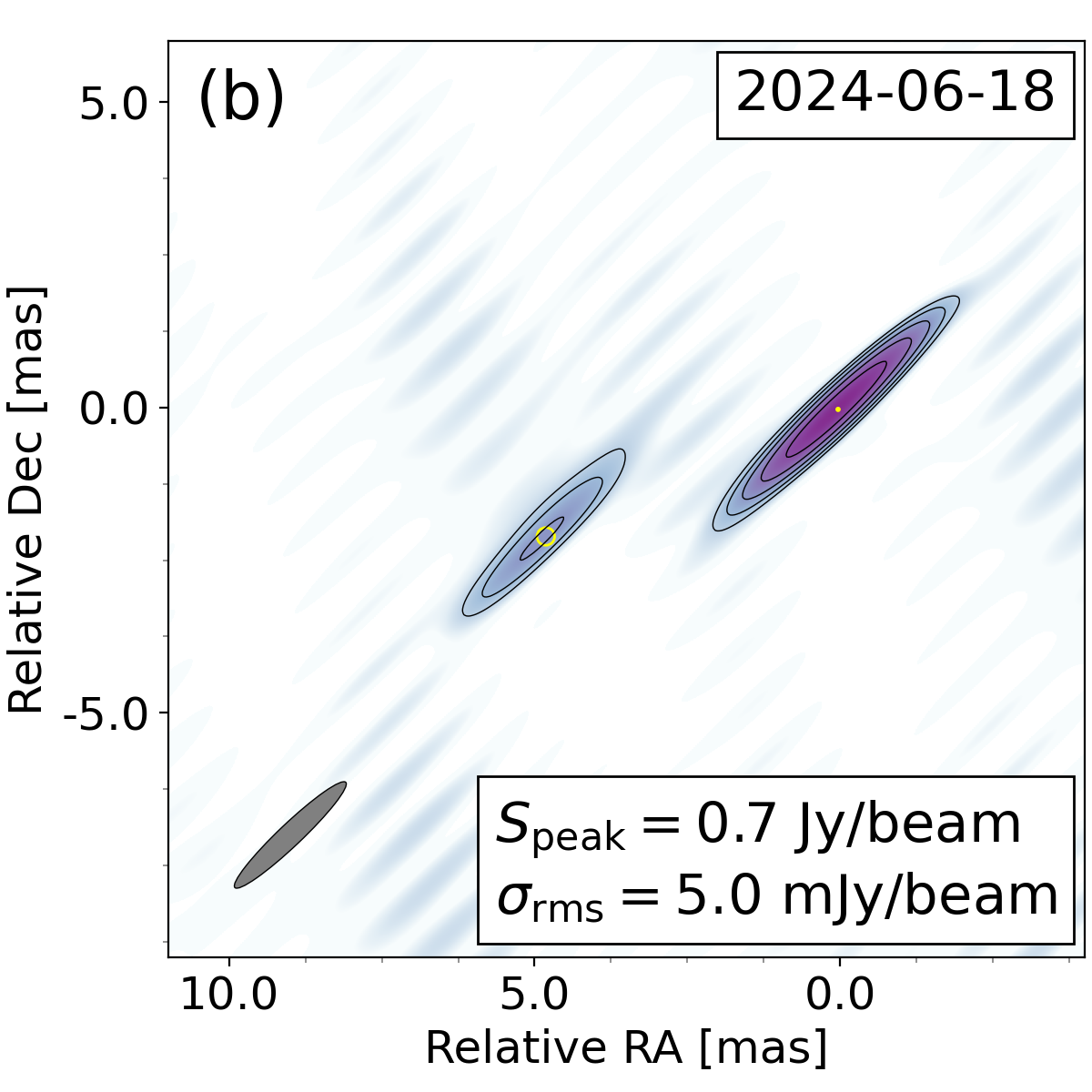}
  \end{subfigure}
  \medskip
  \begin{subfigure}{4.5cm}
    \centering
    \includegraphics[width=4.5cm]{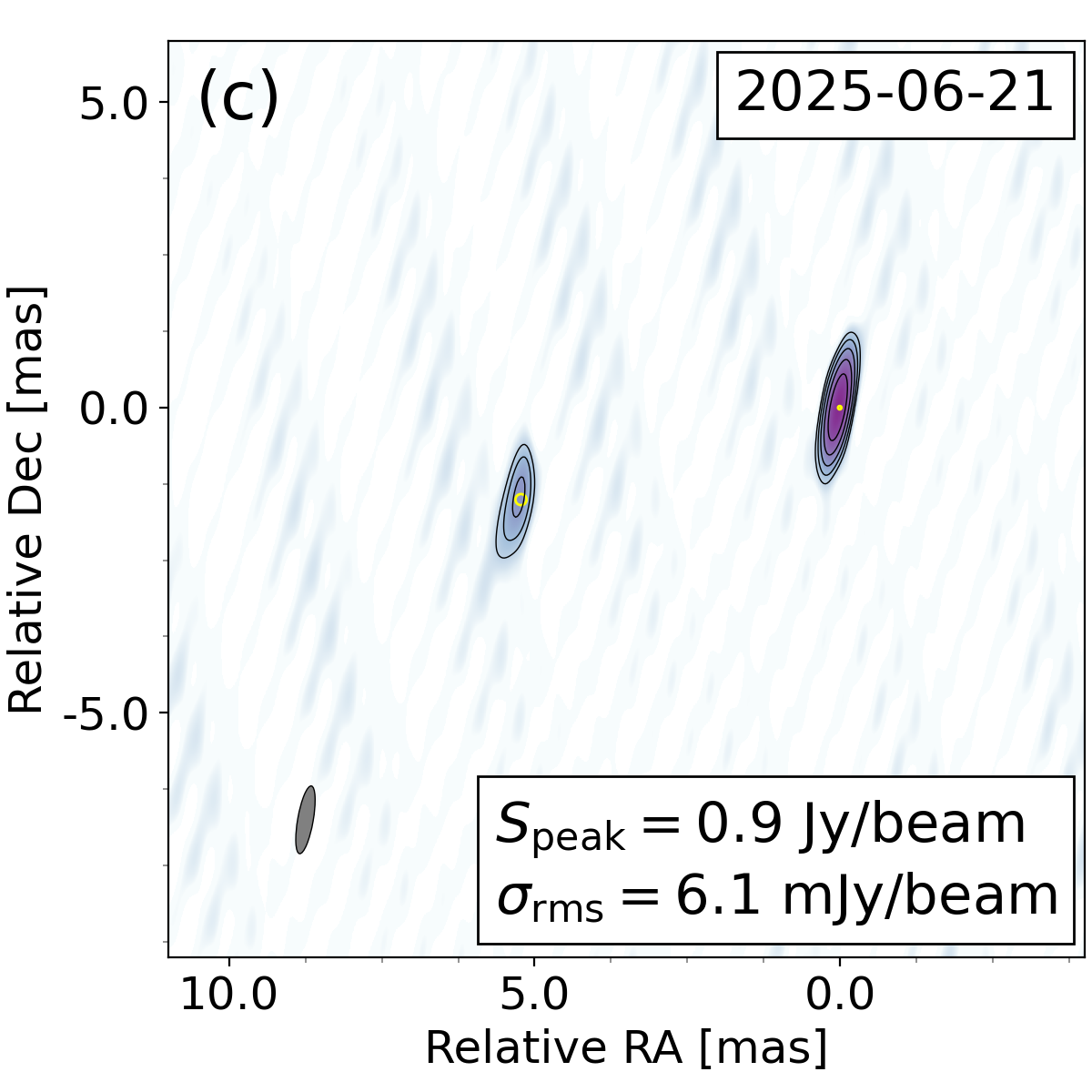}
  \end{subfigure}\hfill
  \begin{subfigure}{4.5cm}
    \centering
  \end{subfigure}
\caption{
{VLBA images of 3C~138 at 23.5 GHz obtained on 2022 December 14 (a), 2024 June 18 (b), and  2025 June 21 (c). Contours are drawn at $(-1,\,1,\,2,\,4,\,\ldots)\times 5\sigma$, where $\sigma$ is the image rms noise. The synthesized beam is shown in the lower-left corner of each panel. Yellow circles denote circular two-dimensional Gaussian model components. Their diameters represent the full widths at half maximum, and their centers mark the component positions.
}
}
\label{fig:vlba_image}
\end{figure}

\begin{figure}[!ht] 
  \centering
    \includegraphics[width=7cm]{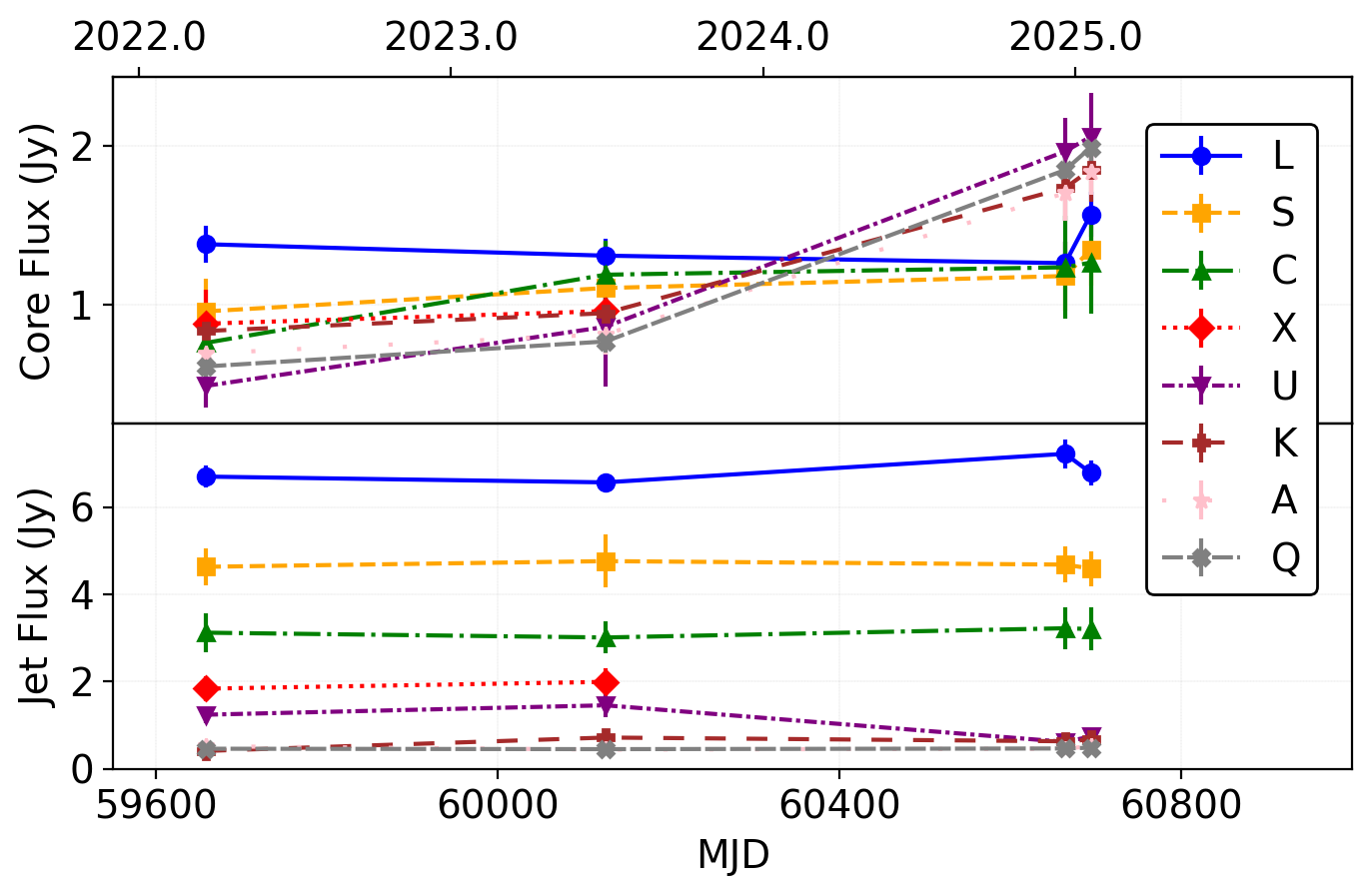}
  \caption{VLA core (top) and jet (bottom) flux densities at 1.5 (L), 3.0 (S), 5.5 (C), 9.0 (X), 14.0 (U), 22.2 (K), 32.0 (A), and 40.1 (Q) GHz as a function of time.}
  \label{fig:VLA_flux_time}
\end{figure}

\begin{figure}[!ht]
\centering
  \begin{subfigure}{4.0cm}
    \centering
    \includegraphics[width=4.0cm]{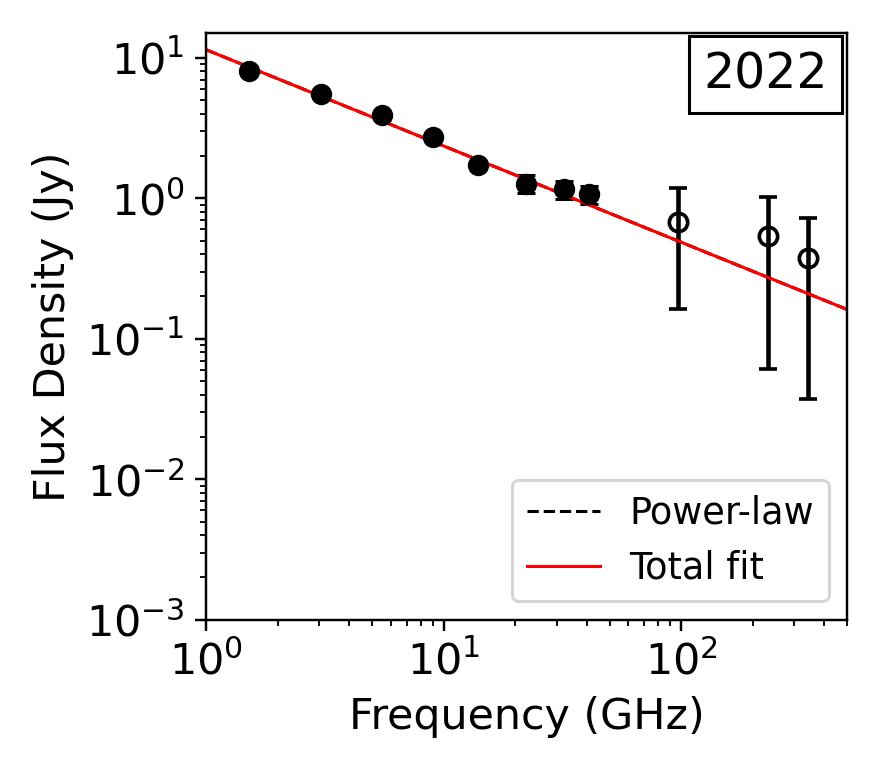}
  \end{subfigure}\hfill
  \begin{subfigure}{4.0cm}
    \centering
    \includegraphics[width=4.0cm]{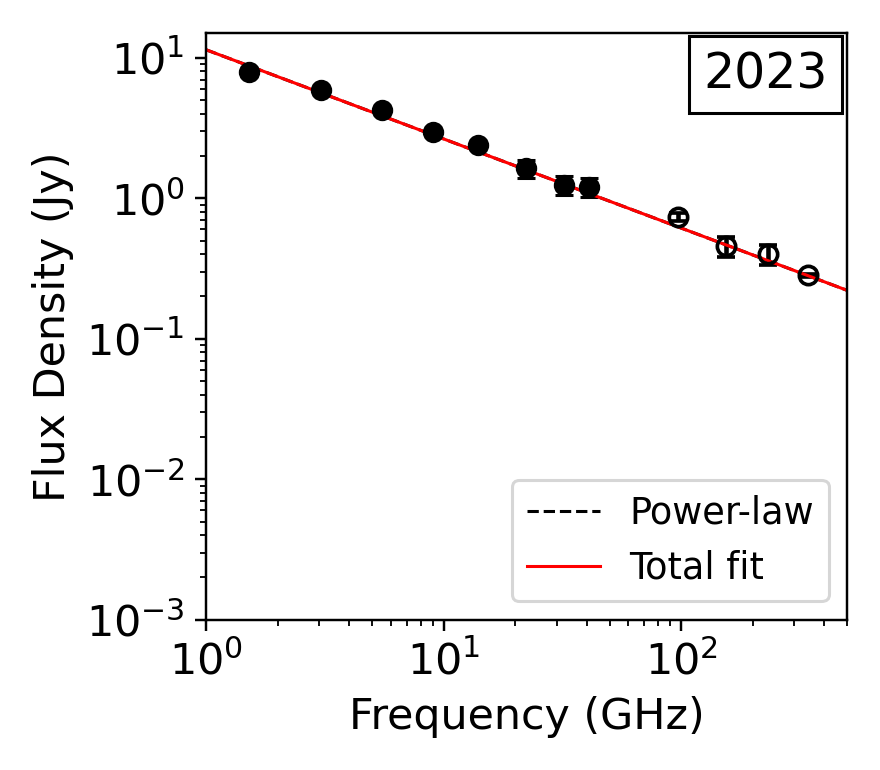}
  \end{subfigure}
  \medskip
  \begin{subfigure}{4.0cm}
    \centering
    \includegraphics[width=4.0cm]{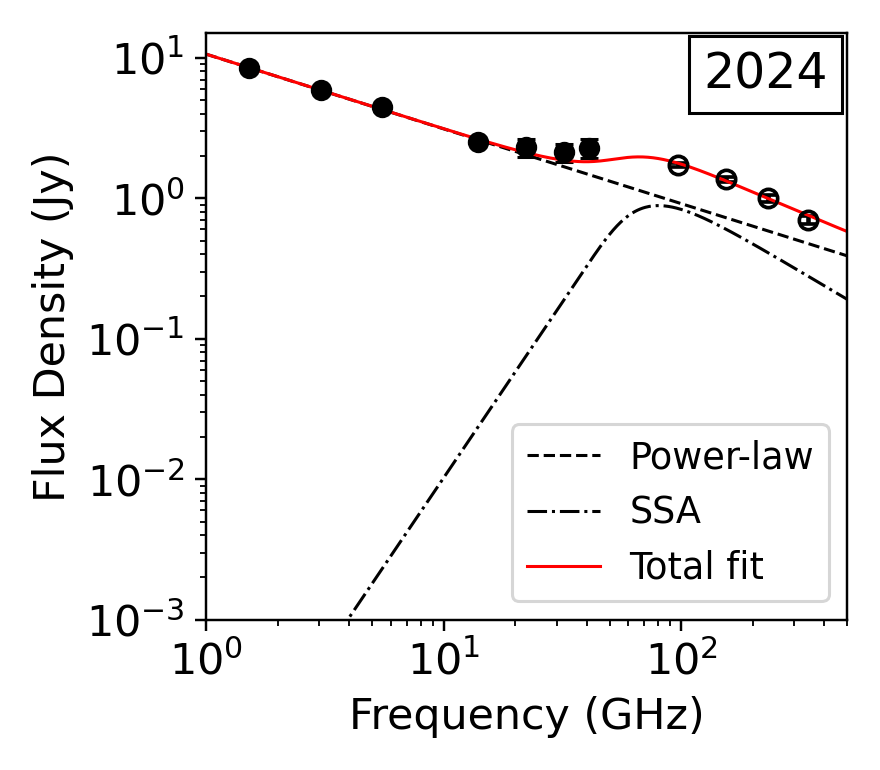}
  \end{subfigure}\hfill
  \begin{subfigure}{4.0cm}
    \centering
    \includegraphics[width=4.0cm]{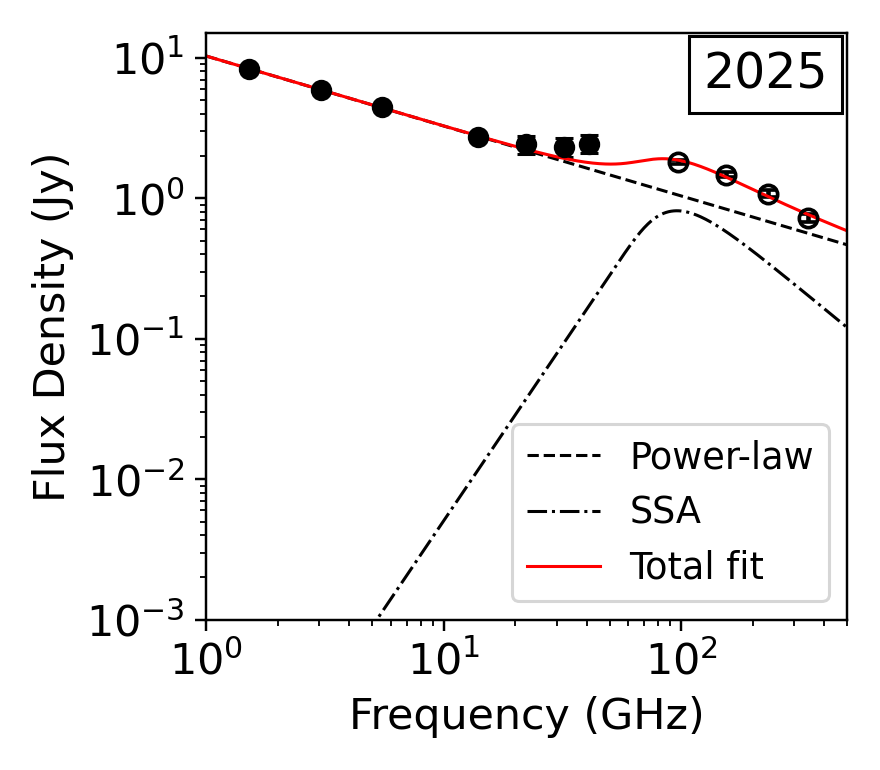}
  \end{subfigure}
\caption{Quasi-simultaneous VLA+ALMA broadband spectra and PL/PL+SSA fits. Solid markers represent VLA observed data, open markers represent GP-interpolated ALMA data (see Sect.~\ref{sec:spectra}).}
\label{fig:vla-alma_fitting}
\end{figure}

\FloatBarrier
\section{Additional tables}
\label{app:tables}

\begin{table}[!ht]
\centering
\footnotesize
\caption{PL+SSA fits to the VLA+ALMA spectra.}
\renewcommand{\arraystretch}{1.25}
\setlength{\tabcolsep}{7pt}
\label{tab:plssa_fits}
\begin{threeparttable}
\begin{tabular}{ccccc}
\hline\hline
Obs. date & $\alpha_{\rm PL}$ & $S_{\rm m}$ & $\nu_{\rm m}$ & $\alpha_0$ \\
(1) & (2) & (3) & (4) & (5) \\
\hline
\rule{0pt}{1.3em}2022-03-20 & $-0.68\pm0.02$ & $\cdots$ & $\cdots$ & $\cdots$  \\
\rule{0pt}{1.3em}2023-07-01 & $-0.63\pm0.01$ & $\cdots$ & $\cdots$ & $\cdots$  \\
\rule{0pt}{1.3em}2024-12-20 & $-0.53\pm0.03$ & $0.62_{-0.25}^{+0.30}$ & $52.88_{-20.19}^{+28.84}$ & $-1.00_{-0.43}^{+0.20}$ \\
\rule{0pt}{1.3em}2025-01-19 & $-0.50\pm0.03$ & $0.63_{-0.24}^{+0.22}$ & $70.61_{-29.53}^{+29.34}$ & $-1.39_{-0.40}^{+0.43}$ \\
[2pt]
\hline
\end{tabular}
\tablefoot{Column designation: (1) Observing date, (2) spectral index of the PL component, (3) peak flux density (Jy) of the SSA component, corresponding (4) turnover frequency (GHz) and (5) optically thin spectral index.}
\end{threeparttable}
\end{table}

\clearpage
\onecolumn

\begin{table}[!ht]
\footnotesize
\caption{Model-fit parameters of the KVN components.}
\centering
\label{tab:kvn_modelfit}
\renewcommand{\arraystretch}{1.1}
\setlength{\tabcolsep}{8pt}
\begin{threeparttable}
\begin{tabular}{cccccccccc}
\hline\hline
Band & Freq & ID & $S_{\rm tot}$ & Radius & $\theta$ & $d$ & $d_{\min}$ & $\sigma_{\rm rms}$ & $S_{\rm peak}$ \\
(1) & (2) & (3) & (4) & (5) & (6) & (7) & (8) & (9) & (10)  \\
\hline
\multicolumn{10}{c}{2024-12-12} \\
\hline
K & 21.801 & W & $1.193 \pm 0.136$ & $\cdots$ & $\cdots$ & $<0.312$ & 0.362 & 7.687 & 1.200 \\
  &        & E & $0.399 \pm 0.059$ & $5.115 \pm 0.047$ & $104.2 \pm 0.5$ & $0.891 \pm 0.094$ & 0.477 & 4.265 & 0.385 \\
Q & 43.346 & W & $1.390 \pm 0.079$ & $\cdots$ & $\cdots$ & $0.277 \pm 0.011$ & 0.095 & 2.242 & 1.367 \\
  &        & E & $0.307 \pm 0.046$ & $5.176 \pm 0.030$ & $104.6 \pm 0.3$ & $0.548 \pm 0.060$ & 0.254 & 3.360 & 0.286 \\
W & 86.436 & W & $1.027 \pm 0.059$ & $\cdots$ & $\cdots$ & $0.051 \pm 0.002$ & 0.046 & 1.671 & 1.025 \\
  &        & E & $0.125 \pm 0.019$ & $5.177 \pm 0.017$ & $104.6 \pm 0.2$ & $0.305 \pm 0.033$ & 0.124 & 1.363 & 0.115 \\
D & 129.398 & W & $0.861 \pm 0.216$ & $\cdots$ & $\cdots$ & $<0.099$ & 0.123 & 26.369 & 0.878 \\
  &         & E & $0.137 \pm 0.037$ & $5.407 \pm 0.021$ & $104.5 \pm 0.2$ & $0.209 \pm 0.043$ & 0.143 & 4.868 & 0.121 \\
\hline
\multicolumn{10}{c}{2025-01-10} \\
\hline
K & 21.801 & W & $1.080 \pm 0.119$ & $\cdots$ & $\cdots$ & $0.455 \pm 0.035$ & 0.391 & 6.468 & 1.081 \\
  &        & E & $0.357 \pm 0.058$ & $5.142 \pm 0.059$ & $103.8 \pm 0.6$ & $1.016 \pm 0.118$ & 0.584 & 4.592 & 0.345 \\
Q & 43.346 & W & $1.237 \pm 0.087$ & $\cdots$ & $\cdots$ & $0.271 \pm 0.014$ & 0.129 & 3.044 & 1.223 \\
  &        & E & $0.233 \pm 0.044$ & $4.948 \pm 0.031$ & $105.9 \pm 0.4$ & $0.459 \pm 0.062$ & 0.348 & 4.033 & 0.226 \\
W & 86.436 & W & $0.980 \pm 0.044$ & $\cdots$ & $\cdots$ & $0.170 \pm 0.005$ & 0.040 & 0.982 & 0.962 \\
  &        & E & $0.156 \pm 0.019$ & $5.333 \pm 0.015$ & $104.7 \pm 0.2$ & $0.346 \pm 0.030$ & 0.108 & 1.095 & 0.144 \\
\hline
\end{tabular}
\tablefoot{
Column designation: 
(1) observing band,
(2) central frequency in GHz,
(3) component identification: W--western component (core), E--eastern component (jet),
(4) model flux density in Jy,
(5) radius in mas,
(6) PA in degrees,
(7) component size in mas,
(8) minimum resolvable size in mas,
(9) uncertainty of post-fit rms in mJy,
(10) peak flux density of the model in Jy beam$^{-1}$.
Uncertainties correspond to 1$\sigma$ errors.
}
\end{threeparttable}
\end{table}

\begin{table}[!ht]
\centering
\footnotesize
\caption{VLBA image parameters.}
\renewcommand{\arraystretch}{1.1}
\setlength{\tabcolsep}{18pt}
\label{tab:vlba_imgparam}
\begin{threeparttable}
\begin{tabular}{ccccccc}
\hline\hline
 Band & Freq & $S_{\rm CLEAN}$ & $S_{\rm p}$ & $\sigma$ & Beam size & Beam PA \\
  (1) & (2) & (3) & (4) & (5) & (6) & (7) \\
\hline
\multicolumn{7}{c}{2022-12-14} \\  
\hline
K & 23.5 & $0.54 \pm 0.08$ & 0.33 & 3.2 & $1.0\times0.6$ & 1.5 \\
\hline
\multicolumn{7}{c}{2024-06-18} \\  
\hline
K & 23.5 & $0.85 \pm 0.13$ & 0.70 & 5.0 & $2.5\times0.4$ & $-46.5$ \\
\hline
\multicolumn{7}{c}{2025-06-21} \\  
\hline
K & 23.5 & $1.18 \pm 0.18$ & 0.95 & 6.1 & $1.1\times0.3$ & $-9.9$ \\
\hline
\end{tabular}
\tablefoot{
Column designation: (1) observing band, (2) central frequency in GHz, (3) total CLEAN flux density with its uncertainty in Jy, (4) peak flux density in Jy\,beam$^{-1}$, (5) image noise level in mJy\,beam$^{-1}$, (6) synthesized beam size in mas $\times$ mas, and (7) beam PA in degrees. 
}
\end{threeparttable}
\end{table}

\begin{table}[!ht]
\footnotesize
\centering
\caption{Model-fit parameters of the VLBA components.}
\renewcommand{\arraystretch}{1.1}
\setlength{\tabcolsep}{8pt}
\label{tab:vlba_modelfit}
\begin{threeparttable}
\begin{tabular}{cccccccccc}
\hline\hline
Band & Freq & ID & $S_{\rm tot}$ & Radius & $\theta$ & $d$ & $d_{\min}$ & $\sigma_{\rm rms}$ & $S_{\rm peak}$ \\
(1) & (2) & (3) & (4) & (5) & (6) & (7) & (8) & (9) & (10)  \\
\hline
\multicolumn{10}{c}{2022-12-14}\\
\hline
K & 23.5 & W & $0.372 \pm 0.052$ & $\cdots$ & $\cdots$ & $0.223 \pm 0.023$ & 0.075 & 3.521 & 0.339 \\
  &      & E & $0.187 \pm 0.041$ & $5.254 \pm 0.027$ & $104.9 \pm 0.3$ & $0.329 \pm 0.055$ & 0.121 & 4.241 & 0.156 \\
\hline
\multicolumn{10}{c}{2024-06-18} \\
\hline
K & 23.5 & W & $0.701 \pm 0.079$ & $\cdots$ & $\cdots$ & $<0.044$ & 0.072 & 4.428 & 0.703 \\
  &      & E & $0.137 \pm 0.037$ & $5.216 \pm 0.031$ & $113.5 \pm 0.3$ & $0.290 \pm 0.062$ & 0.192 & 4.615 & 0.105 \\
\hline
\multicolumn{10}{c}{2025-06-21} \\
\hline
K & 23.5 & W & $0.960 \pm 0.144$ & $\cdots$ & $\cdots$ & $0.056 \pm 0.006$ & 0.054 & 10.706 & 0.936 \\
  &      & E & $0.207 \pm 0.073$ & $5.430 \pm 0.025$ & $106.1 \pm 0.3$ & $0.181 \pm 0.050$ & 0.137 & 11.682 & 0.164 \\
\hline

\end{tabular}
\tablefoot{
Column designation: 
(1) observing band,
(2) central frequency in GHz,
(3) component identification: W--western component (core), E--eastern component (jet),
(4) model flux density in Jy,
(5) radius in mas,
(6) PA in degrees,
(7) component size in mas,
(8) minimum resolvable size in mas,
(9) uncertainty of post-fit rms in mJy,
(10) peak flux density of the model in Jy beam$^{-1}$.
Uncertainties correspond to 1$\sigma$ errors.
}
\end{threeparttable}
\end{table}

\begin{table}[!ht]
\centering
\footnotesize
\caption{VLA image parameters.}
\renewcommand{\arraystretch}{1.25}
\setlength{\tabcolsep}{18pt}
\label{tab:vla_imgparam}
\begin{threeparttable}
\begin{tabular}{cccccccc}
\hline\hline
 Band & Freq & $S_{\rm CLEAN}$ & $S_{\rm p}$ & $\sigma$ & Beam size & Beam PA \\
  (1) & (2) & (3) & (4) & (5) & (6) & (7) \\
\hline
\multicolumn{7}{c}{2022-03-20} \\  
\hline
L & 1.517  & $8.04 \pm 0.40$ & 7.88 & 6.1 & $1766.1\times1143.1$ & 65.6 \\
S & 3.045  & $5.56 \pm 0.28$ & 5.32 & 10.4 & $958.7\times649.2$ & 74.9 \\
C & 5.499  & $3.88 \pm 0.19$ & 3.59 & 9.4 & $728.4\times430.4$ & 76.4 \\
X & 8.999  & $2.71 \pm 0.14$ & 1.92 & 12.6 & $391.5\times229.6$ & 57.3 \\
U & 13.999 & $1.71 \pm 0.09$ & 1.21 & 9.7 & $271.0\times144.5$ & 56.8 \\
K & 22.199 & $1.27 \pm 0.19$ & 0.85 & 4.8 & $161.7\times92.3$  & 59.5 \\
A & 32.019 & $1.16 \pm 0.17$ & 0.68 & 5.3 & $109.6\times53.0$  & 72.1 \\
Q & 40.999 & $1.07 \pm 0.16$ & 0.62 & 1.8 & $66.8\times49.6$   & 51.1 \\
\hline
\multicolumn{7}{c}{2023-07-01} \\  
\hline
L & 1.517  & $7.85 \pm 0.39$ & 7.70 & 4.7 & $2132.8\times1160.7$ & 43.7 \\
S & 3.045  & $5.92 \pm 0.30$ & 5.39 & 8.5 & $855.4\times496.7$ & 49.4 \\
C & 5.499  & $4.22 \pm 0.21$ & 3.45 & 9.9 & $546.9\times382.3$ & 59.9 \\
X & 8.999  & $2.95 \pm 0.15$ & 2.08 & 13.2 & $393.5\times235.2$ & 56.1 \\
U & 13.999 & $2.39 \pm 0.12$ & 1.21 & 12.0 & $146.2\times118.4$ & 55.3 \\
K & 22.199 & $1.64 \pm 0.25$ & 0.95 & 6.2 & $128.4\times91.2$ & 60.6 \\
A & 32.019 & $1.25 \pm 0.19$ & 0.82 & 4.7 & $132.3\times56.2$ & 50.0 \\
Q & 40.999 & $1.21 \pm 0.18$ & 0.77 & 1.7 & $67.9\times52.7$ & 70.0 \\
\hline
\multicolumn{7}{c}{2024-12-20} \\  
\hline
L & 1.517  & $8.49 \pm 0.42$ & 8.31 & 6.9 & $2011.2\times1534.9$ & 19.3 \\
S & 3.045  & $5.86 \pm 0.29$ & 5.56 & 10.0 & $983.3\times702.7$ & 66.1 \\
C & 5.499  & $4.43 \pm 0.22$ & 3.91 & 16.8 & $656.3\times410.3$ & 65.9 \\
U & 13.999 & $2.53 \pm 0.13$ & 2.04 & 13.0 & $222.3\times152.9$ & 58.2 \\
K & 22.199 & $2.30 \pm 0.35$ & 1.76 & 6.5 & $121.2\times94.9$  & 53.4 \\
A & 32.019 & $2.13 \pm 0.32$ & 1.72 & 4.6 & $109.0\times68.7$  & 35.3 \\
Q & 40.999 & $2.29 \pm 0.34$ & 1.84 & 1.9 & $69.7\times49.9$   & 48.9 \\
\hline
\multicolumn{7}{c}{2025-01-19} \\  
\hline
L & 1.517  & $8.38 \pm 0.42$ & 8.14 & 5.7 & $1836.9\times1388.9$ & 39.5 \\
S & 3.045  & $5.90 \pm 0.29$ & 5.52 & 13.8 & $894.0\times652.6$ & 74.3 \\
C & 5.499  & $4.44 \pm 0.22$ & 3.97 & 18.2 & $684.6\times408.0$ & 73.8 \\
U & 13.999 & $2.72 \pm 0.14$ & 2.14 & 15.7 & $207.7\times140.9$ & 55.2 \\
K & 22.199 & $2.43 \pm 0.36$ & 1.87 & 5.3 & $115.0\times91.2$  & 52.2 \\
A & 32.019 & $2.33 \pm 0.35$ & 1.84 & 3.7 & $97.8\times69.2$   & 43.6 \\
Q & 40.999 & $2.45 \pm 0.37$ & 1.99 & 1.9 & $70.7\times52.5$   & 49.5 \\
\hline
\end{tabular}
\tablefoot{
Column designation: (1) observing band, (2) central frequency in GHz, (3) total CLEAN flux density with its uncertainty in Jy, (4) peak flux density in Jy\,beam$^{-1}$, (5) image noise level in mJy\,beam$^{-1}$, (6) synthesized beam size in mas $\times$ mas, and (7) beam PA in degrees. 
}
\end{threeparttable}
\end{table}

\longtab[6]{
\begingroup
\centering
\footnotesize
\renewcommand{\arraystretch}{1.25}
\setlength{\tabcolsep}{6pt}
\begin{longtable}{cccccccccc}
\caption{Model-fit parameters of the VLA components.}\label{tab:vla_modelfit}\\
\hline\hline
Band & Freq & ID & $S_{\rm tot}$ & Radius & $\theta$ & $d$ & $d_{\min}$ & $\sigma_{\rm rms}$ & $S_{\rm peak}$ \\
(1) & (2) & (3) & (4) & (5) & (6) & (7) & (8) & (9) & (10)  \\
\hline
\endfirsthead
\caption{continued.}\\
\hline\hline
Band & Freq & ID & $S_{\rm tot}$ & Radius & $\theta$ & $d$ & $d_{\min}$ & $\sigma_{\rm rms}$ & $S_{\rm peak}$ \\
(1) & (2) & (3) & (4) & (5) & (6) & (7) & (8) & (9) & (10)  \\
\hline
\endhead
\endfoot
\multicolumn{10}{c}{2022-03-20} \\
\hline
L & 1.517  & C  & $1.379 \pm 0.116$ & $\cdots$ & $\cdots$ & $157.265 \pm 9.401$ & 79.712  & 4.836  & 1.358 \\
  &        & J1 & $6.702 \pm 0.248$ & $424.641 \pm 0.550$ & $72.5 \pm 0.6$ & $43.968 \pm 1.151$  & 34.929  & 4.570  & 6.675 \\
S & 3.045  & C  & $0.955 \pm 0.206$ & $\cdots$ & $\cdots$ & $<53.458$           & 111.574 & 21.784 & 0.972 \\
  &        & J1 & $4.632 \pm 0.421$ & $366.827 \pm 1.161$ & $68.4 \pm 1.8$ & $<35.843$           & 47.549  & 19.013 & 4.629 \\
C & 5.499  & C  & $0.756 \pm 0.221$ & $\cdots$ & $\cdots$ & $<64.237$           & 105.705 & 31.051 & 0.785 \\
  &        & J1 & $3.120 \pm 0.440$ & $350.680 \pm 2.664$ & $67.2 \pm 8.7$ & $51.971 \pm 5.166$  & 52.158  & 30.777 & 3.146 \\
X & 8.999  & C  & $0.878 \pm 0.213$ & $\cdots$ & $\cdots$ & $<28.064$           & 48.256  & 25.152 & 0.869 \\
  &        & J1 & $1.836 \pm 0.287$ & $347.039 \pm 6.684$ & $71.1 \pm 44.4$ & $<23.519$          & 31.180  & 22.220 & 1.824 \\
U & 13.999 & C  & $0.485 \pm 0.138$ & $\cdots$ & $\cdots$ & $<12.089$           & 36.628  & 18.780 & 0.493 \\
  &        & J1 & $1.238 \pm 0.183$ & $360.562 \pm 9.640$ & $72.1 \pm 18.8$ & $30.538 \pm 3.245$  & 19.701  & 13.434 & 1.203 \\
K & 22.199 & C  & $0.832 \pm 0.104$ & $\cdots$ & $\cdots$ & $<8.360$            & 10.109  & 6.474  & 0.837 \\
  &        & J1 & $0.135 \pm 0.036$ & $269.355 \pm 1.716$ & $74.3 \pm 0.4$ & $<18.048$           & 21.603  & 4.701  & 0.135 \\
  &        & J2 & $0.272 \pm 0.056$ & $381.957 \pm 0.788$ & $72.0 \pm 0.1$ & $<10.917$           & 16.512  & 5.697  & 0.278 \\
A & 32.019 & C  & $0.689 \pm 0.094$ & $\cdots$ & $\cdots$ & $11.092 \pm 1.083$  & 6.974   & 6.331  & 0.671 \\
  &        & J1 & $0.123 \pm 0.032$ & $236.346 \pm 4.127$ & $74.5 \pm 1.0$ & $38.249 \pm 8.076$  & 14.967  & 3.937  & 0.092 \\
  &        & J2 & $0.122 \pm 0.023$ & $321.323 \pm 1.654$ & $73.1 \pm 0.3$ & $23.332 \pm 3.294$  & 10.063  & 2.127  & 0.109 \\
  &        & J3 & $0.269 \pm 0.040$ & $396.663 \pm 0.628$ & $71.2 \pm 0.2$ & $19.736 \pm 2.186$  & 7.909   & 2.992  & 0.247 \\
Q & 40.999 & C  & $0.608 \pm 0.079$ & $\cdots$ & $\cdots$ & $4.937 \pm 0.453$   & 4.945   & 5.108  & 0.613 \\
  &        & J1 & $0.096 \pm 0.029$ & $247.835 \pm 4.204$ & $75.1 \pm 1.0$ & $33.798 \pm 8.364$  & 13.199  & 4.051  & 0.070 \\
  &        & J2 & $0.091 \pm 0.025$ & $329.493 \pm 3.397$ & $72.5 \pm 0.6$ & $30.434 \pm 6.784$  & 11.917  & 3.285  & 0.069 \\
  &        & J3 & $0.271 \pm 0.040$ & $399.449 \pm 1.141$ & $70.9 \pm 0.2$ & $24.654 \pm 2.824$  & 6.171   & 2.931  & 0.226 \\
\hline
\multicolumn{10}{c}{2023-07-01} \\
\hline
L & 1.517  & C  & $1.307 \pm 0.105$ & $\cdots$ & $\cdots$ & $146.387 \pm 8.288$ & 84.274 & 4.197  & 1.294 \\
  &        & J1 & $6.569 \pm 0.213$ & $418.896 \pm 0.295$ & $71.0 \pm 0.1$ & $<16.084$ & 33.896 & 3.450  & 6.564 \\
S & 3.045  & C  & $1.103 \pm 0.294$ & $\cdots$ & $\cdots$ & $<69.970$ & 112.513 & 37.829 & 1.141 \\
  &        & J1 & $4.762 \pm 0.608$ & $364.967 \pm 1.358$ & $78.9 \pm 1.6$ & $<50.271$ & 55.040 & 38.558 & 4.797 \\
C & 5.499  & C  & $1.187 \pm 0.214$ & $\cdots$ & $\cdots$ & $69.590 \pm 13.536$ & 55.422 & 19.020 & 1.152 \\
  &        & J1 & $3.010 \pm 0.366$ & $347.586 \pm 0.966$ & $71.7 \pm 0.9$ & $34.358 \pm 1.931$ & 36.837 & 22.145 & 3.022 \\
X & 8.999  & C  & $0.957 \pm 0.245$ & $\cdots$ & $\cdots$ & $<31.509$ & 51.723 & 30.496 & 0.946 \\
  &        & J1 & $1.992 \pm 0.318$ & $350.508 \pm 1.229$ & $71.6 \pm 1.1$ & $<26.023$ & 32.275 & 25.059 & 1.977 \\
U & 13.999 & C  & $0.858 \pm 0.376$ & $\cdots$ & $\cdots$ & $<9.655$ & 34.807 & 75.697 & 0.993 \\
  &        & J1 & $0.242 \pm 0.152$ & $252.998 \pm 14.622$ & $78.0 \pm 3.1$ & $<5.963$ & 47.405 & 40.926 & 0.299 \\
  &        & J2 & $1.212 \pm 0.232$ & $377.124 \pm 1.930$ & $71.3 \pm 0.3$ & $19.669 \pm 3.861$ & 16.845 & 21.804 & 1.185 \\
K & 22.199 & C  & $0.944 \pm 0.202$ & $\cdots$ & $\cdots$ & $14.347 \pm 0.643$ & 15.295 & 21.083 & 0.942 \\
  &        & J1 & $0.122 \pm 0.036$ & $257.973 \pm 4.053$ & $75.8 \pm 0.9$ & $<15.303$ & 21.156 & 5.174  & 0.122 \\
  &        & J2 & $0.592 \pm 0.181$ & $387.826 \pm 7.551$ & $71.1 \pm 1.1$ & $49.298 \pm 15.102$ & 24.166 & 25.588 & 0.466 \\
A & 32.019 & C  & $0.811 \pm 0.130$ & $\cdots$ & $\cdots$ & $<2.726$ & 9.134 & 10.257 & 0.812 \\
  &        & J1 & $0.126 \pm 0.041$ & $256.830 \pm 10.446$ & $75.2 \pm 2.3$ & $33.232 \pm 20.892$ & 20.068 & 6.221 & 0.105 \\
  &        & J2 & $0.312 \pm 0.055$ & $389.829 \pm 2.774$ & $71.6 \pm 0.4$ & $21.840 \pm 5.548$ & 10.570 & 4.819 & 0.285 \\
Q & 40.999 & C  & $0.766 \pm 0.071$ & $\cdots$ & $\cdots$ & $<0.883$ & 3.692 & 3.295 & 0.767 \\
  &        & J1 & $0.095 \pm 0.031$ & $249.468 \pm 2.253$ & $75.2 \pm 0.5$ & $39.052 \pm 4.507$ & 14.735 & 4.465 & 0.067 \\
  &        & J2 & $0.089 \pm 0.024$ & $330.397 \pm 3.104$ & $72.7 \pm 0.5$ & $32.745 \pm 6.208$ & 11.973 & 3.015 & 0.068 \\
  &        & J3 & $0.263 \pm 0.035$ & $398.352 \pm 1.353$ & $70.9 \pm 0.2$ & $26.807 \pm 2.705$ & 5.726 & 2.255 & 0.219 \\
\hline
\multicolumn{10}{c}{2024-12-20} \\
\hline
L & 1.517  & C  & $1.259 \pm 0.135$ & $\cdots$ & $\cdots$ & $<23.176$ & 125.127 & 7.186  & 1.254 \\
  &        & J1 & $7.225 \pm 0.335$ & $401.121 \pm 0.695$ & $78.8 \pm 0.8$ & $<42.375$ & 54.154  & 7.751  & 7.204 \\
S & 3.045  & C  & $1.178 \pm 0.216$ & $\cdots$ & $\cdots$ & $<17.502$ & 100.495 & 19.438 & 1.183 \\
  &        & J1 & $4.683 \pm 0.415$ & $348.229 \pm 0.994$ & $74.8 \pm 1.2$ & $<31.675$ & 48.964  & 18.360 & 4.679 \\
C & 5.499  & C  & $1.233 \pm 0.325$ & $\cdots$ & $\cdots$ & $<30.871$ & 90.224  & 41.469 & 1.231 \\
  &        & J1 & $3.222 \pm 0.484$ & $332.155 \pm 0.889$ & $72.7 \pm 1.6$ & $<16.719$ & 51.717  & 36.025 & 3.219 \\
U & 13.999 & C  & $1.964 \pm 0.214$ & $\cdots$ & $\cdots$ & $<4.936$  & 13.286  & 11.545 & 1.968 \\
  &        & J1 & $0.614 \pm 0.147$ & $377.257 \pm 6.327$ & $72.7 \pm 1.0$ & $70.005 \pm 12.653$ & 31.069 & 16.935 & 0.535 \\
K & 22.199 & C  & $1.733 \pm 0.208$ & $\cdots$ & $\cdots$ & $<5.745$  & 8.512   & 12.376 & 1.740 \\
  &        & J1 & $0.173 \pm 0.047$ & $275.669 \pm 5.490$ & $75.2 \pm 1.1$ & $52.056 \pm 10.979$ & 21.039 & 5.965  & 0.140 \\
  &        & J2 & $0.456 \pm 0.064$ & $391.105 \pm 1.684$ & $71.8 \pm 0.2$ & $32.454 \pm 3.368$  & 10.428 & 4.439  & 0.417 \\
A & 32.019 & C  & $1.701 \pm 0.173$ & $\cdots$ & $\cdots$ & $5.703 \pm 0.410$   & 5.840   & 8.800  & 1.710 \\
  &        & J1 & $0.097 \pm 0.025$ & $242.484 \pm 3.244$ & $75.9 \pm 0.8$ & $33.531 \pm 6.488$  & 15.594 & 2.985  & 0.083 \\
  &        & J2 & $0.083 \pm 0.018$ & $320.675 \pm 1.651$ & $74.0 \pm 0.3$ & $20.521 \pm 3.301$  & 12.997 & 1.954  & 0.077 \\
  &        & J3 & $0.279 \pm 0.028$ & $397.476 \pm 0.592$ & $71.3 \pm 0.1$ & $16.521 \pm 1.184$  & 5.819  & 1.362  & 0.266 \\
Q & 40.999 & C  & $1.847 \pm 0.116$ & $\cdots$ & $\cdots$ & $4.399 \pm 0.196$   & 2.471   & 3.652  & 1.838 \\
  &        & J1 & $0.107 \pm 0.035$ & $249.870 \pm 4.578$ & $75.2 \pm 1.0$ & $34.659 \pm 9.156$  & 14.425 & 5.143  & 0.079 \\
  &        & J2 & $0.087 \pm 0.026$ & $329.035 \pm 3.914$ & $72.8 \pm 0.7$ & $32.686 \pm 7.828$  & 13.108 & 3.576  & 0.066 \\
  &        & J3 & $0.270 \pm 0.044$ & $399.143 \pm 1.526$ & $71.2 \pm 0.2$ & $24.677 \pm 3.051$  & 6.828  & 3.424  & 0.227 \\
\hline
\multicolumn{10}{c}{2025-01-19} \\
\hline
L & 1.517  & C  & $1.567 \pm 0.141$ & $\cdots$ & $\cdots$ & $200.732 \pm 12.836$ & 95.861 & 6.283  & 1.543 \\
  &        & J1 & $6.787 \pm 0.287$ & $399.674 \pm 0.663$ & $70.9 \pm 0.7$ & $44.313 \pm 1.327$ & 44.914 & 6.068  & 6.777 \\
S & 3.045  & C  & $1.345 \pm 0.227$ & $\cdots$ & $\cdots$ & $101.664 \pm 12.266$ & 86.271 & 18.941 & 1.320 \\
  &        & J1 & $4.591 \pm 0.403$ & $359.055 \pm 1.042$ & $70.4 \pm 1.3$ & $<33.578$ & 44.496 & 17.588 & 4.583 \\
C & 5.499  & C  & $1.263 \pm 0.325$ & $\cdots$ & $\cdots$ & $<25.820$ & 89.287 & 40.428 & 1.270 \\
  &        & J1 & $3.203 \pm 0.495$ & $324.215 \pm 1.883$ & $70.1 \pm 2.6$ & $<34.475$ & 54.087 & 37.839 & 3.207 \\
U & 13.999 & C  & $2.057 \pm 0.274$ & $\cdots$ & $\cdots$ & $<8.029$ & 15.101 & 18.142 & 2.064 \\
  &        & J1 & $0.738 \pm 0.151$ & $374.915 \pm 5.568$ & $72.9 \pm 0.9$ & $71.330 \pm 11.136$ & 24.946 & 14.891 & 0.626 \\
K & 22.199 & C  & $1.850 \pm 0.207$ & $\cdots$ & $\cdots$ & $<3.357$ & 7.578 & 11.468 & 1.854 \\
  &        & J1 & $0.191 \pm 0.050$ & $270.025 \pm 6.151$ & $75.2 \pm 1.3$ & $58.473 \pm 12.302$ & 20.040 & 6.096 & 0.144 \\
  &        & J2 & $0.477 \pm 0.060$ & $391.899 \pm 1.738$ & $71.6 \pm 0.3$ & $36.674 \pm 3.475$ & 9.097 & 3.739 & 0.420 \\
A & 32.019 & C  & $1.834 \pm 0.142$ & $\cdots$ & $\cdots$ & $<3.930$ & 4.224 & 5.482 & 1.838 \\
  &        & J1 & $0.108 \pm 0.029$ & $239.620 \pm 2.803$ & $76.2 \pm 0.7$ & $28.137 \pm 5.606$ & 15.261 & 3.640 & 0.095 \\
  &        & J2 & $0.103 \pm 0.020$ & $319.812 \pm 1.871$ & $73.2 \pm 0.3$ & $26.100 \pm 3.742$ & 11.026 & 1.863 & 0.092 \\
  &        & J3 & $0.306 \pm 0.032$ & $395.174 \pm 0.840$ & $71.2 \pm 0.1$ & $21.686 \pm 1.680$ & 5.978 & 1.692 & 0.284 \\
Q & 40.999 & C  & $1.989 \pm 0.125$ & $\cdots$ & $\cdots$ & $3.653 \pm 0.162$ & 2.540 & 3.904 & 1.984 \\
  &        & J1 & $0.038 \pm 0.011$ & $209.919 \pm 1.780$ & $78.7 \pm 0.5$ & $17.505 \pm 3.559$ & 11.529 & 1.395 & 0.035 \\
  &        & J2 & $0.080 \pm 0.017$ & $261.175 \pm 1.876$ & $74.4 \pm 0.4$ & $23.103 \pm 3.752$ & 9.237 & 1.797 & 0.070 \\
  &        & J3 & $0.092 \pm 0.023$ & $332.260 \pm 3.211$ & $73.0 \pm 0.6$ & $32.066 \pm 6.423$ & 11.360 & 2.747 & 0.071 \\
  &        & J4 & $0.271 \pm 0.039$ & $401.075 \pm 1.353$ & $71.0 \pm 0.2$ & $24.696 \pm 2.706$ & 6.252 & 2.756 & 0.232 \\
\hline
\end{longtable}
\tablefoot{
Column designation: 
(1) observing band,
(2) central frequency in GHz,
(3) component identification: W--western component (core), E--eastern component (jet),
(4) model flux density in Jy,
(5) radius in mas,
(6) PA in degrees,
(7) component size in mas,
(8) minimum resolvable size in mas,
(9) uncertainty of post-fit rms in mJy,
(10) peak flux density of the model in Jy beam$^{-1}$.
Uncertainties correspond to 1$\sigma$ errors.
}
\endgroup
}
\end{appendix}
\end{document}